\documentclass[12pt,a4paper]{article}%
\pdfoutput=1

\usepackage[T1]{fontenc}
\usepackage[utf8]{inputenc}
\usepackage{amsmath}
\usepackage{amssymb}
\usepackage{amsthm}
\usepackage[ 
    ]{hyperref}
\usepackage{mathtools}
\usepackage{graphicx}
\usepackage{booktabs}
\usepackage{caption}
\usepackage{slashed}
\usepackage{listings}
\usepackage{slashed}
\usepackage{rotating}
\usepackage{feynmp}
\usepackage{subfigure}
\usepackage{geometry}
\usepackage{multicol}
\usepackage{dsfont}
\usepackage[nottoc]{tocbibind}
\usepackage{multicol}
\usepackage{authblk}
\usepackage{cite}
\usepackage{appendix}
\usepackage{color}

\numberwithin{equation}{section} 
\setcaptionmargin{1cm}

\title{\LARGE \bf Gravitational Waves from Axion Monodromy}
\author{Arthur Hebecker}
\author{Joerg Jaeckel}
\author{Fabrizio Rompineve}
\author{Lukas T. Witkowski}
\date{June 24, 2016}

\affil{Institute for Theoretical Physics, University of Heidelberg, Philosophenweg 19, 69120 Heidelberg, Germany}

\begin{document}

\maketitle

\begin{abstract}

\noindent Large field inflation is arguably the simplest and most natural variant of slow-roll inflation. Axion monodromy may be the most promising framework for realising this scenario. As one of its defining features, the long-range polynomial potential possesses short-range, instantonic modulations. These can give rise to a series of local minima in the post-inflationary region of the potential. We show that for certain parameter choices the inflaton populates more than one of these vacua inside a single Hubble patch. This corresponds to a dynamical phase decomposition, analogously to what happens in the course of thermal first-order phase transitions. In the subsequent process of bubble wall collisions, the lowest-lying axionic minimum eventually takes over all space. Our main result is that this violent process sources gravitational waves, very much like in the case of a first-order phase transition. We compute the energy density and peak frequency of the signal, which can lie anywhere in the mHz-GHz range, possibly within reach of next-generation interferometers. We also note that this "dynamical phase decomposition" phenomenon and its gravitational wave signal are more general and may apply to other inflationary or reheating scenarios with axions and modulated potentials.

\end{abstract}

\newpage

\begin{section}{Introduction}

The central predictions allowing us to discriminate between inflationary models are the slow roll parameters, most prominently the tilt of the scalar power spectrum $n_{s}\simeq 0.96$ and the tensor-to-scalar ratio $r\lesssim 0.08$ \cite{Ade:2015xua, Ade:2015tva}. However, even with growing precision many models are expected to remain consistent with this limited set of data. Hence it is of great importance to identify additional predictions characterizing specific classes of models. 

Here, our main focus will be axion monodromy inflation \cite{Silverstein:2008sg, McAllister:2008hb} (see \cite{Marchesano:2014mla, Blumenhagen:2014gta, Hebecker:2014eua} for its supergravity incarnation as `$\!F$-term axion monodromy' and \cite{ Blumenhagen:2014nba, Hebecker:2014kva, Ibanez:2014swa, Escobar:2015fda, Kobayashi:2015aaa, Hebecker:2015tzo} for a selection of related recent work). In this promising class of models, the inflaton originally enjoys a discrete shift symmetry, explicitly broken by a polynomial term, such as $m^2\phi^2$. The aim of this paper is to describe a new and potentially striking observational signature which is peculiar to axion monodromy inflation. For previous work on the phenomenology of such inflationary models see \cite{Peiris:2013opa, Easther:2013kla, Kobayashi:2014ooa, Flauger:2014ana, Flauger:2016idt, Ashoorioon:2008pj} (see also \cite{Parameswaran:2016qqq, Kadota:2016jlw} for closely related potentials) and, in particular, \cite{Moghaddam:2015ava, Adshead:2015pva, McDonough:2016xvu} for preheating and \cite{Amin:2011hj} for oscillon dynamics in this context (see in particular \cite{Zhou:2013tsa} for gravitational radiation from preheating in oscillon models).

In short, our message is the following: Due to the typical instantonic modulations of the potential, first order phase-transition-like, violent dynamics may occur after the end of inflation and before reheating. This leads to additional gravitational waves, which are of course very different in frequency from those studied in the CMB. Thus, monodromy models may (in addition to or independently of their prediction of $r$) be established by future ground- or space-based interferometers. The final word would thus come from the new field of gravitational-wave astronomy \cite{Abbott:2016blz} (for a recent review see e.g.~\cite{Blair:2016idv}).

Of course, gravitational waves are a well-known signature of cosmic strings, including axionic strings. For recent work on the non-trival late-time dynamics of axionic models and gravitational waves which is closer in spirit to our proposal see, e.g. \cite{Hiramatsu:2012sc, Daido:2015gqa, Daido:2015bva, Higaki:2016yqk, Higaki:2016jjh, Ashoorioon:2015hya}. Emission of gravitational waves from axionic couplings in inflationary setups has recently been considered in \cite{Domcke:2016bkh, Peloso:2016gqs} (see also \cite{Ferreira:2014zia, Ferreira:2015omg} for constraints). Independently of the gravitational wave signal, related dynamical phenomena may also occur in the dark matter context \cite{D'Amico:2016kqm, Jaeckel:2016qjp}.

Let us now explain the physics underlying our scenario in more detail: The basic building block of monodromy inflation is an axion with sub-planckian decay constant $f$. Non-perturbative effects induce the familiar $\cos(\phi/f)$-type potential. When the monodromy effect is included, this cosine-potential shows up in the form of modulations of the long-range, polynomial term. Even if the relative size of these modulations is small at large field values, where slow-roll inflation is realized, they can become dominant near the minimum of the polynomial potential. After inflation, the field oscillates with decreasing amplitude such that its motion eventually becomes confined to the vicinity of one of these local minima. However, due to field fluctuations, different local minima may be chosen in different regions of the same Hubble patch. In other words, the Universe is decomposed into phases.

Two comments are in order. First, the field fluctuations inducing the above phenomenon can have different origin. On the one hand, there are inflationary super-horizon fluctuations which have become classical at the time when they re-enter the horizon. On the other hand, the field is subject to an intrinsic quantum uncertainty at any given time. As we will see, this second type of uncertainty is, after parametric amplification \cite{Jaeckel:2016qjp}, more likely to source the desired phase decomposition.

Second, while we for definiteness identify the monodromic axion with the inflaton, the phenomenon can occur also in different contexts. In particular, similar physics may arise in any non-inflationary axion model with a monodromy (the relaxion being one recent popular example \cite{Graham:2015cka}, see also \cite{Ibanez:2015fcv}). Moreover, even within the inflationary context, our proposal of `dynamical phase decomposition' is not restricted to monodromy models. Indeed, models of `aligned' or `winding' inflation \cite{Kim:2004rp, Berg:2009tg, Ben-Dayan:2014zsa} can naturally exhibit short-range modulations on top of a long-range periodic potential. The Weak Gravity Conjecture \cite{ArkaniHamed:2006dz} for instantons may in fact demand such modulations \cite{Rudelius:2015xta, Brown:2015iha} and the simplest string constructions naturally provide them \cite{Hebecker:2015rya}. Furthermore, the Weak Gravity Conjecture for domain walls constrains models based on monodromy \cite{Ibanez:2015fcv, Hebecker:2015zss}. In particular, the size of the wiggles is bounded \cite{Hebecker:2015zss}.

Let us now complete the discussion of the cosmological dynamics: After a phase decomposition has occurred, the regions with the lowest-lying populated minimum will expand. This is very similar to the way in which a strong first-order phase transition is completed through the collision of cosmic bubbles of true vacuum. However, in our case the transition occurs before reheating and very far from thermal equilibrium. It may thus be more appropriate to talk about `dynamical phase decomposition' rather than about a phase transition in the usual sense. Nevertheless, the concept of bubble formation and collision is still appropriate in our setting.

Thermodynamic cosmological phase transitions have been widely explored in various contexts (see \cite{Witten:1984rs, Kosowsky:1992rz, Kosowsky:1991ua} for early seminal work and \cite{Grojean:2006bp} and refs. therein for the case of the electroweak phase transition). In particular, it is well known that they source gravitational radiation (see \cite{Hindmarsh:2015qjv, Caprini:2015zlo, Guzzetti:2016mkm} for recent reviews and e.g. \cite{Buchmuller:2013lra} for radiation from other cosmological sources such as cosmic strings and preheating). We will rely on these results. 

This paper is structured as follows: Sec.~\ref{sec:axionmonodromy} provides the basic setup. More specifically, Sec.~\ref{sub:phasesmon} introduces our axion monodromy setting with dominant quadratic potential and a series of local minima, Sec.~\ref{sub:reheating} briefly discusses reheating, and Sec.~\ref{sub:energydensity} explains the post-inflationary dynamics and the phase decomposition. In Sec.~\ref{sec:phases} we estimate the probability that a phase decomposition occurs as a consequence of field fluctuations. In particular, we treat inflationary fluctuations in Sec.~\ref{sub:evolution} and the intrinsic quantum uncertainty in Sec.~\ref{sub:quantumflucts}. In Sec.~\ref{sub:numerics} we address the crucial issue of the possible enhancements of fluctuations due to background-field-oscillations in the modulated potential. Ultimately, in Sec.~\ref{sec:gravitationalwaves} we estimate the spectrum and abundance of the gravitational radiation produced during the phase transition before we conclude in Sec.~\ref{sec:conclusions}. Additionally, we devote Appendix~\ref{sec:after} to a more detailed discussion of scalar field fluctuations after inflation. Appendix~\ref{sec:inflection} provides some details of an inflection point model of inflation in which our gravitational wave signal may also arise.

\end{section}

\begin{section}{Phases from axion monodromy}
\label{sec:axionmonodromy}

In this section we introduce a string-motivated scenario of the inflationary universe. It is based on the framework of Axion Monodromy Inflation \cite{Silverstein:2008sg, McAllister:2008hb}. We begin by explaining the basic features of the axion potential and the possibility of having coexisting populated axionic vacua. 

\begin{subsection}{Local minima in axion monodromy}
\label{sub:phasesmon}

The inflaton potential of a model of axion monodromy inflation contains an oscillatory term, which respects a discrete shift symmetry, and a polynomial term, which breaks it explicitly. In this work we will take the polynomial term to be quadratic, as this will be sufficient to demonstrate the effect we wish to study:
\begin{equation}
\label{eq:potential}
V(\phi)=\frac{1}{2}m^{2}\phi^{2}+\Lambda^4\cos\left(\frac{\phi}{f}+\gamma\right).
\end{equation}
The potential \eqref{eq:potential}, plotted in Fig.~\ref{fig:potential1}, can have local minima, whose existence depends on the values of the prefactor $\Lambda^4$, the so-called axion decay constant $f$ and the inflaton mass $m$.  Although we have in mind the specific case in which $\phi$ is the inflaton, many of our considerations apply to the case of a generic axion-like field with potential \eqref{eq:potential}.\footnote{While we expect a similar qualitative behaviour the numerical results may depart significantly from the values we find here.}

We begin with an analysis of the classical evolution of $\phi$. The equation of motion of $\phi$ reads:
\begin{equation}
\label{eq:eom}
\ddot{\phi}+3H\dot{\phi}+V'(\phi)=0,
\end{equation}
where the prime denotes a derivative with respect to $\phi$. For constant Hubble rate $H$ and temporarily neglecting the cosine term in \eqref{eq:potential}, the solution of \eqref{eq:eom} is
\begin{equation}
\phi\sim e^{-\frac{3}{2}Ht}\cos(\omega t).
\end{equation}
Therefore, the amplitude of $\phi$ decreases. This conclusion remains valid even for time-varying $H$. In fact, since $H(t)\sim \rho_{\phi}^{1/2}/M_{p}$, the Hubble rate decreases as the amplitude of $\phi$ falls. Once the amplitude is sufficiently small, the cosine oscillations in \eqref{eq:potential} cannot be neglected any longer. Eventually, the field is caught in one of the cosine wells. One can observe that the field is more likely to get trapped in one of the lowest-lying minima. This can be understood as follows: the friction term in \eqref{eq:eom} becomes less and less relevant as $H(t)$ falls. This implies that the fractional energy loss per oscillation also decreases. Therefore, even though the field can in principle get stuck at any time, it is more likely to do so late in its evolution, when it is oscillating near the bottom of the well containing the lowest-lying minima.\footnote{The actual argument to show that wells at the bottom of the potential are more likely to host the inflaton requires more care. In Sec.~\ref{sub:numerics} we provide numerical examples that support this statement.}  For this reason, we will mostly focus on the last two wells. 

The existence of different local minima implies that the universe is potentially decomposed into several phases. This happens if the field settles in different minima in different parts of the universe. Due to fluctuations, the scalar $\phi$ can end up in one or the other minimum in different regions of the same Hubble patch. In this paper, we are interested in studying the conditions under which such a phase decomposition can take place. 

If such a phenomenon occurs, eventually the field will settle in the state of lower energy as a consequence of the expansion of bubbles containing the true vacuum. This corresponds to a phase transition, which can in principle have strong cosmological signatures, above all the radiation of gravitational waves. We would like to provide an estimate for the spectrum of gravitational waves produced in such an event (see also \cite{Hiramatsu:2012sc, Daido:2015gqa, Daido:2015bva, Higaki:2016yqk, Higaki:2016jjh} for related work, but in different contexts).

Before moving on to study the details of this scenario, let us determine the condition on the parameters $\Lambda, f, m$ for the potential to exhibit local minima. 
In order to have local minima, the equation $V'=0$ must have non-vanishing solutions. Let us, for a moment, simplify by setting $\gamma=0$. We then have:
\begin{equation}
V'=0\qquad \Rightarrow \qquad m^{2}\phi=\frac{\Lambda^4}{f}\sin\left(\frac{\phi}{f}\right).
\end{equation}
Graphically, it is clear that this equation has non-vanishing solutions only if
\begin{equation}
\label{eq:condition}
\kappa\equiv \frac{\Lambda^4}{f^2 m^{2}} \geq 1.
\end{equation}
Here we have used $\gamma=0$, but the equation remains parametrically valid even for $\gamma\neq 0$.
Under this condition the potential has the form represented in Fig.~(\ref{fig:potential1}). 
\begin{figure}
    \centering
    \includegraphics[width=0.5\textwidth]{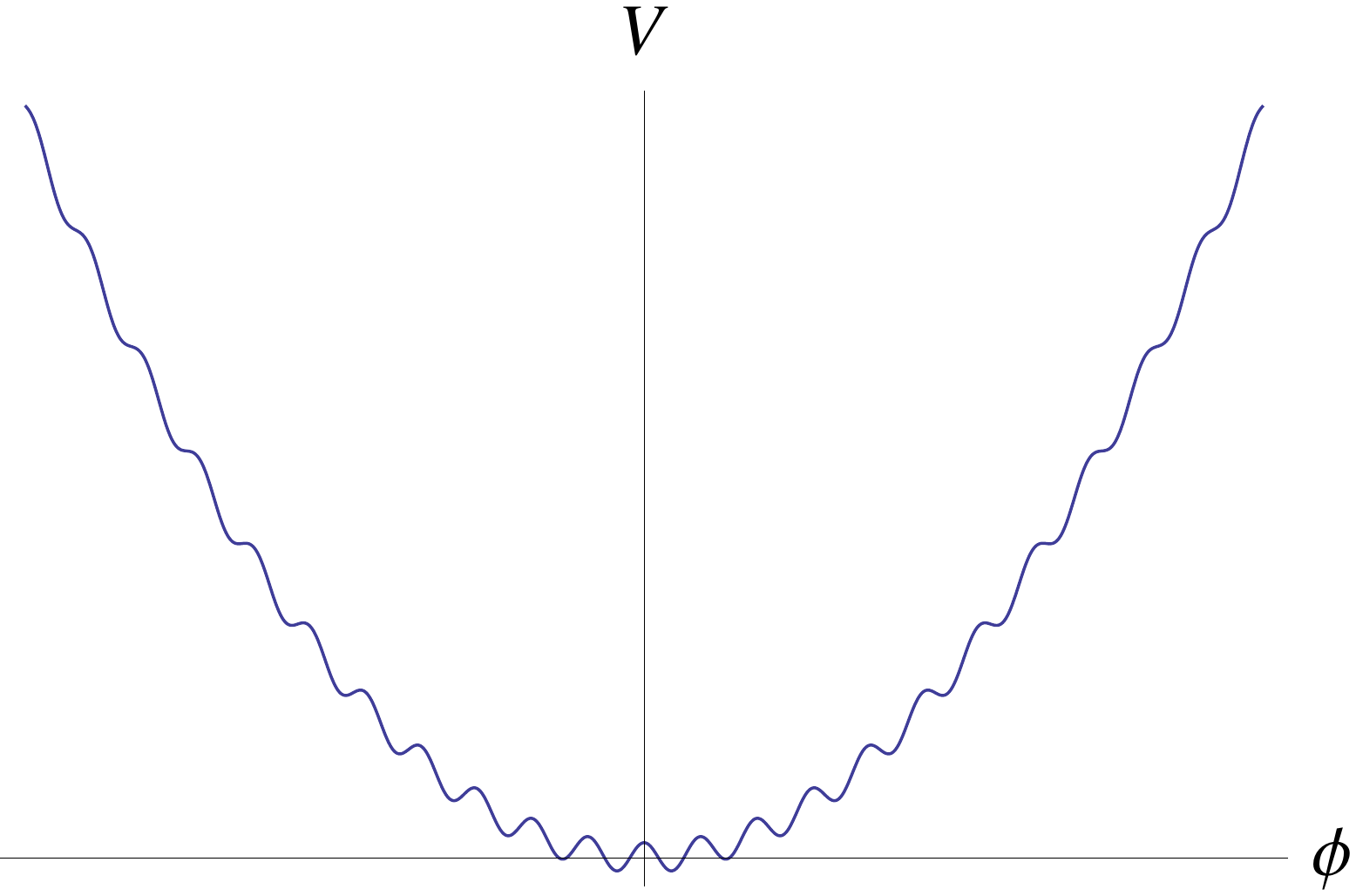}
    \caption{Monodromy potential, as in \eqref{eq:potential} with $\gamma=0$. The parameter $\kappa/\pi$ approximately measures the number of local minima. Here $\kappa\simeq 50$.}
    \label{fig:potential1}
\end{figure}
Practically however, as we have already remarked, we will focus only on the two lowest local minima, which are in general non-degenerate for $\gamma\neq 0$ (see Fig.~\ref{fig:dw2}).\footnote{The choice $\gamma=0$ may lead to stable domain walls, which are generically problematic.}
\begin{figure}
    \centering
    \includegraphics[width=0.5\textwidth]{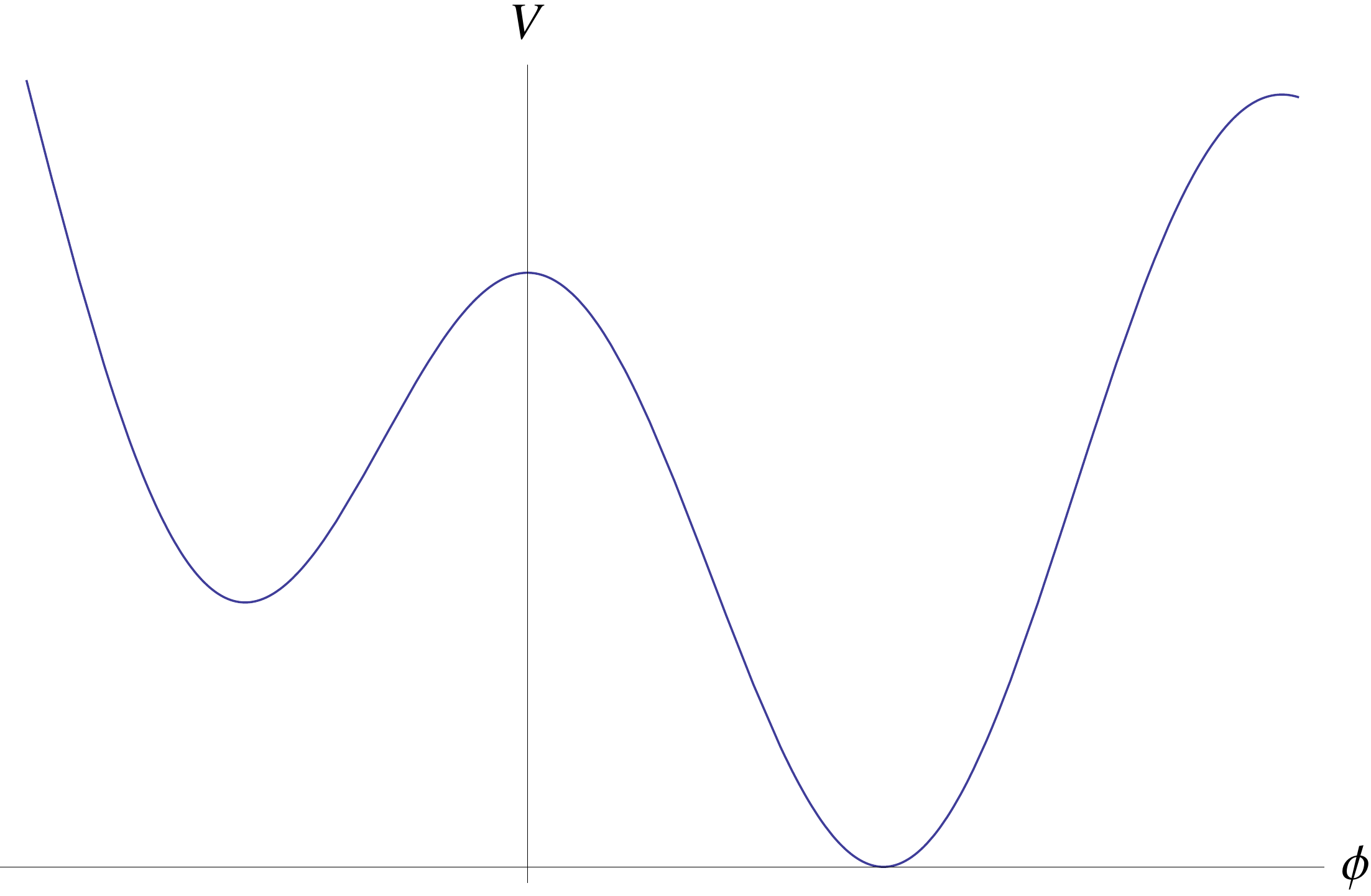}
    \caption{Non-degenerate two-well potential, as it can be obtained from \eqref{eq:potential} by focusing on the two wells closest to the origin.}
    \label{fig:dw2}
\end{figure}

We are now ready to move on to a detailed discussion of phase decomposition in our scenario. However, before doing so, a few comments about reheating are in order.
\end{subsection}

\begin{subsection}{Reheating}
\label{sub:reheating}

Typically, after inflation the Universe undergoes a so-called \emph{reheating} phase, where the energy density in the inflaton sector is transferred to standard model degrees of freedom, and possibly to dark sectors. Our analysis of the dynamics of the inflaton after inflation needs to take this into account. We therefore first focus on the following question: does reheating happen before or after the inflaton field is caught in one of the cosine wells?

In order to answer this question, we need to specify the interactions of the inflaton with matter and/or radiation. Here let us consider the example of a Planck-suppressed modulus-like coupling to a scalar field $\chi$. The largest decay rate, barring the possibility of parametric resonances, is obtained if $\chi$ enjoys a coupling $(1/M_{p})\chi^2 \Box \phi$, leading to 
\begin{equation}
\label{eq:decayrate}
\Gamma_{\phi\rightarrow \chi\chi}\sim \frac{m_{\phi}^{3}}{M_{p}^2}.
\end{equation}
This can arise if $\chi$ is a Higgs boson (see e.g. \cite{Cicoli:2012aq, Higaki:2012ar}). Decay rates to gauge bosons may also be of this type. The inflaton may also couple to scalar matter as $(1/M_{p})\phi\bar{\chi}\Box\chi+c.c.$, but this leads to even smaller decay rates. 

The strategy is now as follows: we assume that the field is trapped in one of the cosine wells, and compare the perturbative decay rate \eqref{eq:decayrate} with the Hubble rate $H^{2}\sim \Lambda^4/M_{p}^2$, as we are assuming that the field is oscillating in the last wells. For the same reason, we should take $m^2_{\phi}= V^{''}|_{min}\simeq\Lambda^4/f^2$ in \eqref{eq:decayrate}. As usual in cosmology, the condition for the decay to be efficient is $\Gamma\gtrsim H$. If we find $\Gamma<H$, then we will be consistent with our assumption that the field is first caught in one of the wells and decays only later. Now,
\begin{equation}
\label{eq:reheat}
\Gamma_{\phi}< H\quad \Rightarrow \quad \kappa\left(\frac{m}{f}\right)\left(\frac{m}{M_p}\right)< 1.
\end{equation}
The inequality \eqref{eq:reheat} is easily satisfied in our setup, as in quadratic inflation $m\sim 10^{-5} M_{p}$ and $f$ may be only slightly smaller than $M_{p}$, while $\kappa\gtrsim O(1)$.  Therefore the field generically decays perturbatively only after getting caught in one of the cosine wells. In what follows, we will therefore not consider the decay of $\phi$ any longer.
\end{subsection}

\begin{subsection}{Field oscillations and damping}
\label{sub:energydensity}

We first focus on how the Hubble parameter changes after inflation. The energy density of the universe is a sum of three terms: $\rho=V_{\phi}+T_{\phi}-V_{\lambda}$, where $V_{\lambda}$ is the energy density due to the cosmological constant. We absorb $V_{\lambda}$ in $V_{\phi}$ and define $V_{\phi}$ such that the global minimum has vanishing potential energy. The evolution of the Hubble parameter and of the scalar field $\phi$ is dictated by the Friedmann equation, together with the equation of motion of $\phi$:
\begin{align}
\label{eq:hevolution}
3H^{2}&=\frac{1}{2}m^{2}\dot{\phi}^{2}+V_{\phi} \\
\label{eq:phievol}
\ddot{\phi}+3H\dot{\phi}&=-V_{\phi}'.
\end{align}
Since the energy density is decreasing due to friction, it is clear that also $H(t)$ will decrease with $t$. However, whenever $\dot{\phi}=0$, $\phi$ is undamped and the energy density is constant. In consequence $H(t)$ is stationary as well. The typical behaviour of $\phi(t)$ and $H(t)$ is shown in Fig.~\ref{fig:evolutions}.
\begin{figure}
\centering
\subfigure{\includegraphics[width=0.45\textwidth]{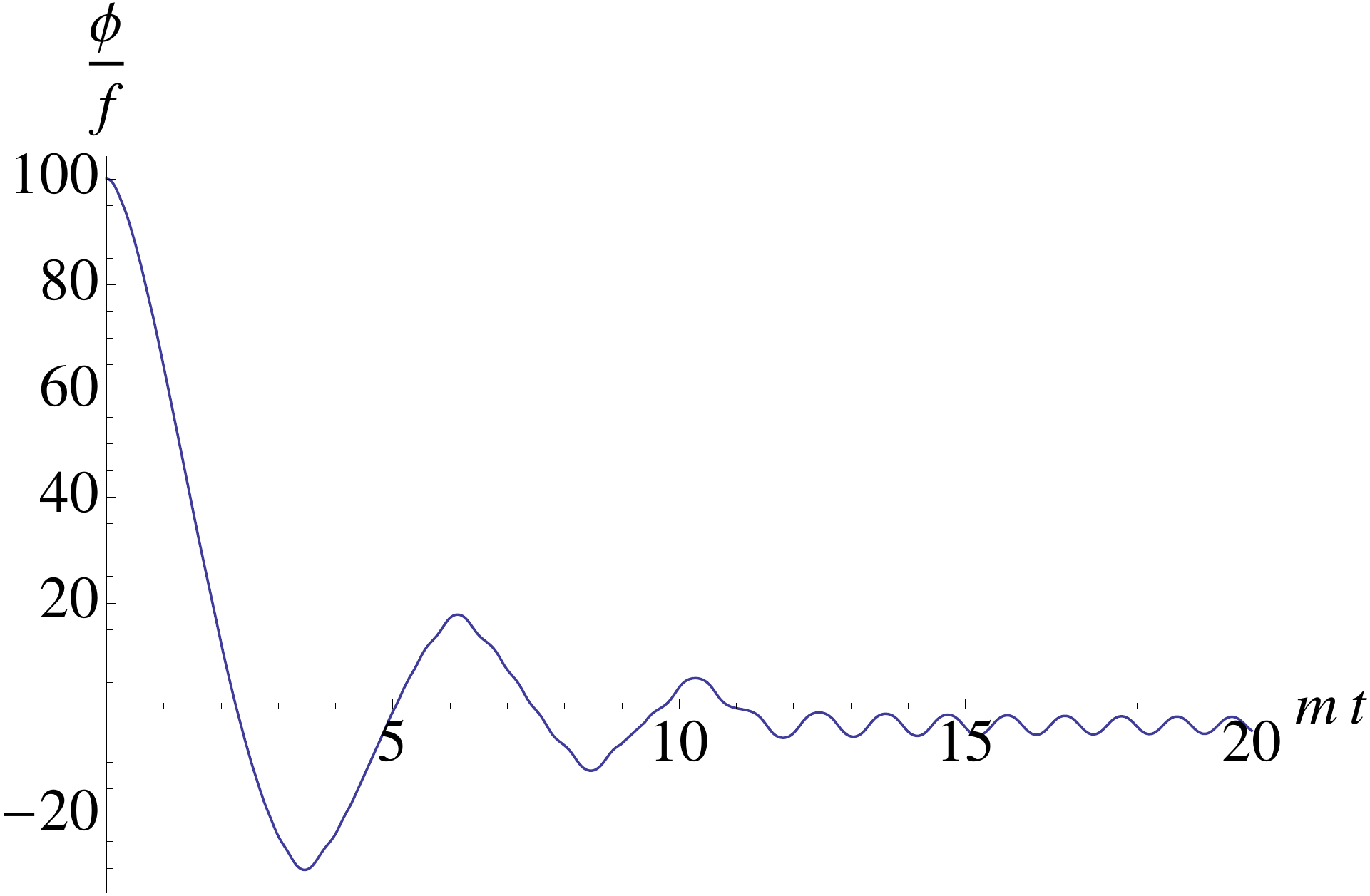}}
\subfigure{\includegraphics[width=0.45\textwidth]{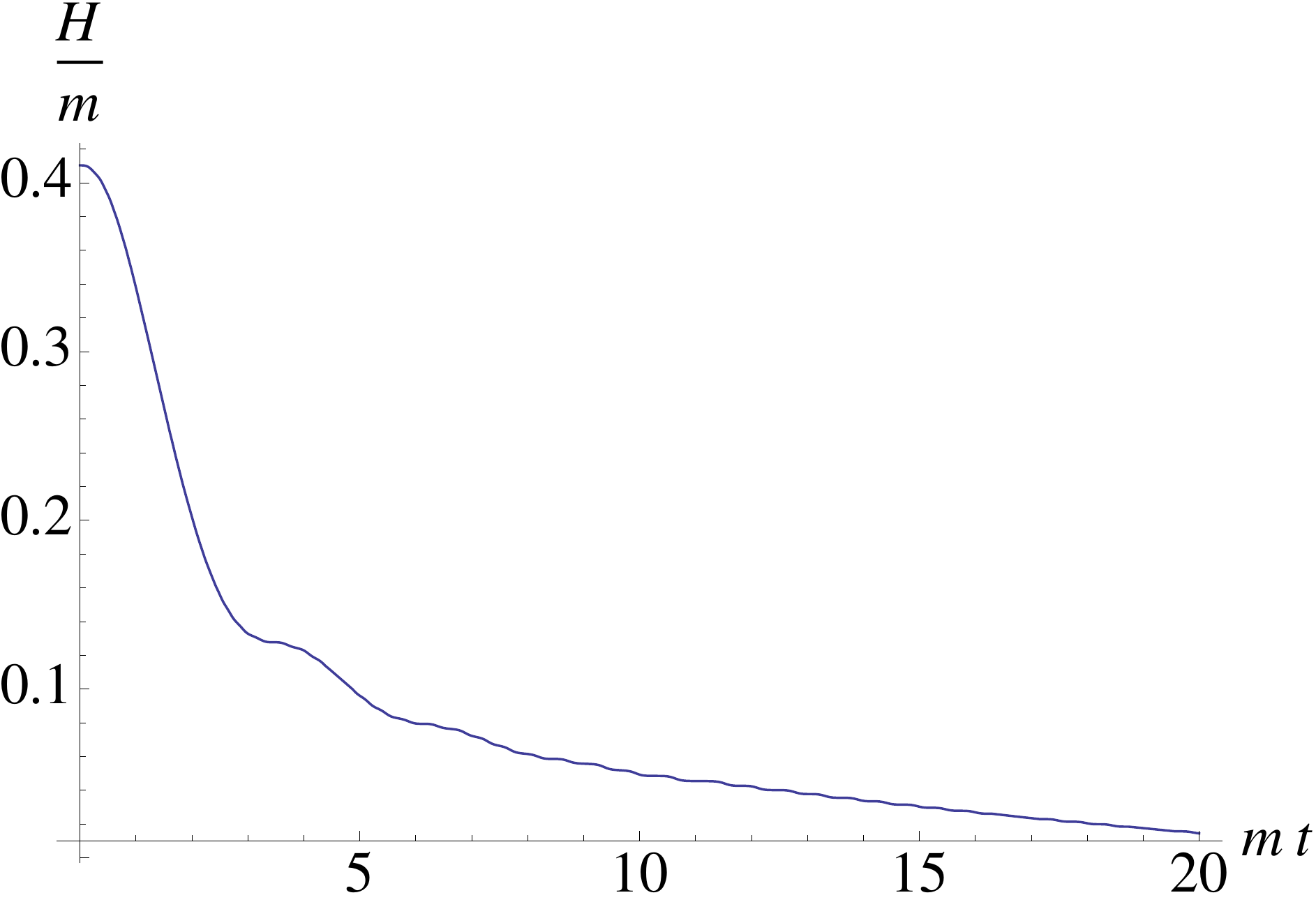}}
\caption{Evolution of: (a) the scalar field $\phi$  and (b) the Hubble rate $H$ according to \eqref{eq:phievol} and \eqref{eq:hevolution}. $V_{\phi}$ is as in \eqref{eq:potential} and $V_{\lambda}=0$. The initial condition is $\phi(t_{0})=M_{p}$. Furthermore we have chosen $\kappa=60, f=10^{-2} M_{p}$. The scalar field oscillates around $\phi=0$ over a wide field range crossing several wells, before getting caught in one of the local minima at $t\approx 11/m$.}
\label{fig:evolutions}
\end{figure}

Let us now examine the energy density in the inflaton field in more detail. Its amplitude decreases after each oscillation due to Hubble friction. At some point the energy density $\rho_{\phi}$ is comparable to the height of the last cosine wells. The field is then caught in one of the local minima, depending on the initial conditions. In the absence of spatial field inhomogeneities, the inflaton will populate only one of these two minima. This situation is shown in Fig.~\ref{fig:energydecrease}.
\begin{figure}
    \centering
    \includegraphics[width=0.5\textwidth]{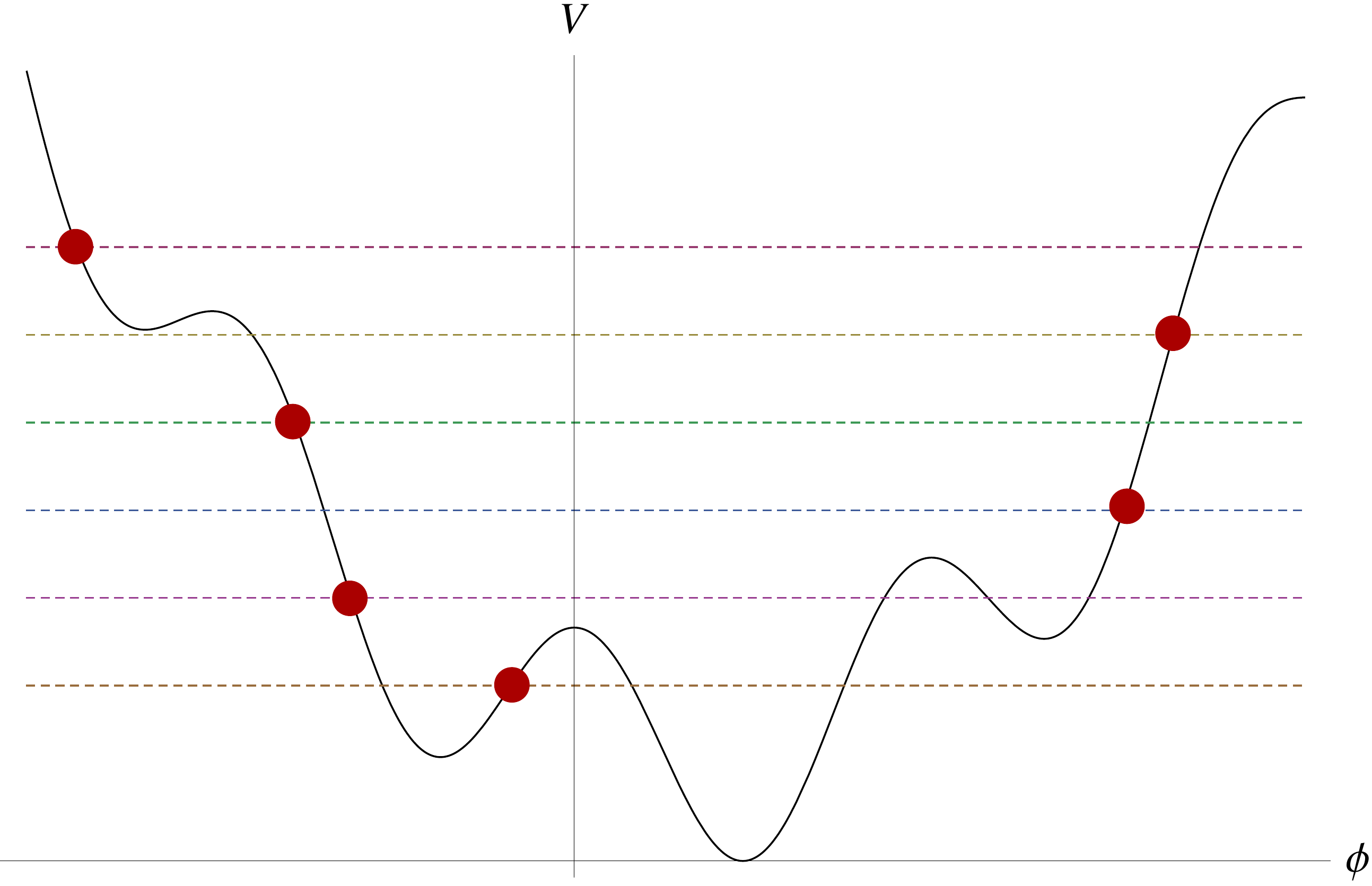}
    \caption{Monodromy potential, as in \eqref{eq:potential}, with $\gamma\neq 0$. The axes are chosen such that the local maximum between the last two wells sits at $\phi=0$ and the lowest minimum has $V=0$. Here $\kappa\approx 12$. The turning points in the field trajectory are shown. The amplitude of the oscillations decreases as a consequence of Hubble friction.}
    \label{fig:energydecrease}
\end{figure}

The conclusions can radically change in the presence of field fluctuations $\delta\phi(t,\mathbf{x})$. In this case, the field may end up in one or the other minimum in different regions belonging to the same Hubble patch. The lines drawn in Fig.~\ref{fig:energydecrease} become bands of a certain width, corresponding to the uncertainty $\delta\rho$ in the energy density induced by $\delta\phi$ (see Fig.~\ref{fig:energybands}). Now suppose that, as a consequence of friction, the energy density has decreased to a value close to the height of the barrier separating the two last minima in Fig.~\ref{fig:energydecrease}. During the next oscillation, the field will start rolling inside one of the two wells, say the one on the left of Fig.~\ref{fig:energybands}. At the end of the oscillation, since its energy density is smaller than the height of the barrier separating the two minima, the field is very likely to remain confined in the left well. However, due to the field fluctuations, there is a non-vanishing probability that the field actually reaches the other well on the right hand side and remains there in some regions of the Universe. If this is the case, at different point in the same Hubble patch the field lives in different minima. Therefore, the Universe decomposes into two phases.
\begin{figure}
    \centering
    \includegraphics[width=0.6\textwidth]{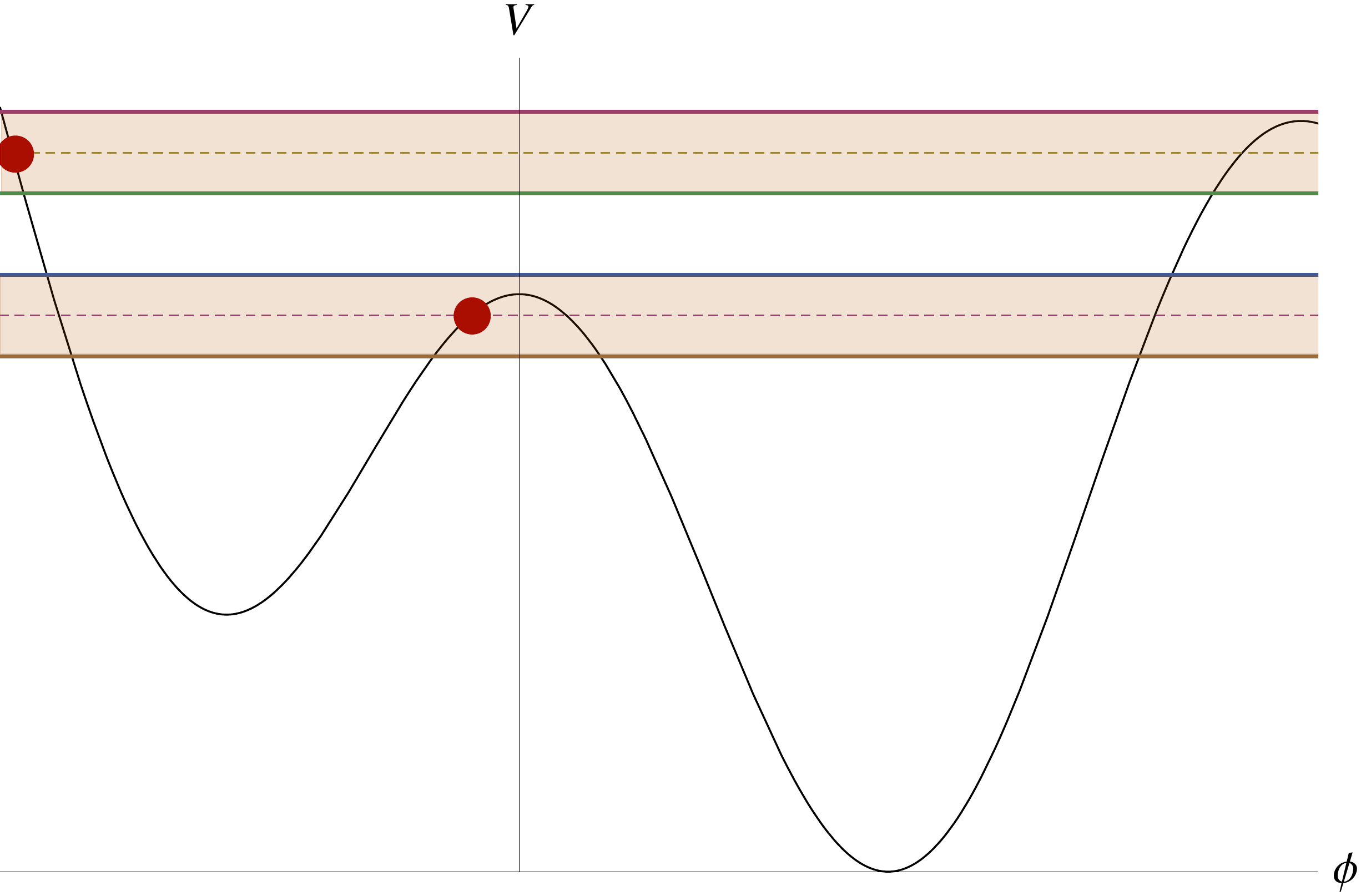}
    \caption{Same as Fig.~\ref{fig:energydecrease}, but now only the last two wells are shown. The width of the shaded bands represent the uncertainty of the scalar field energy density $\delta\rho$, due to field fluctuations. The distance between the two bands corresponds to the energy loss $\Delta\rho$ due to friction.}
    \label{fig:energybands}
\end{figure}
This is precisely the situation we are interested in.

Quantitatively, let us compute the energy density lost during one oscillation, as a consequence of Hubble friction. In general this is not an easy task given the complicated shape of the potential in Fig.~\ref{fig:energydecrease}. However, it is greatly simplified by focusing only on the last cosine wells. In this case, the loss of energy during a half oscillation inside a single well can be estimated by a quadratic approximation of the potential:
\begin{equation}
V_{approx}=\frac{1}{2}M^{2}\phi^{2},
\end{equation}
with $M^2=|V^{''}|_{min}$, i.e. the curvature of the potential \eqref{eq:potential} around the minimum of the well, where $\cos(\phi/f)\approx -1$. We obtain
\begin{equation}
M^2=m^{2}+\frac{\Lambda^4}{f^2}=(1+\kappa)m^2
\end{equation}
since we are interested in the regime $\kappa>1$, we can take $M^2\approx \Lambda^4/f^2$.
Immediately after inflation the Hubble rate evolves approximately as during matter domination, i.e.
\begin{equation}
H=\frac{2}{3t};\qquad \rho\sim a^{-3}\sim t^{-2},
\end{equation}
so that the relative decrease in energy density in one half period $\Delta t\sim M^{-1}/2$ is given by
\begin{equation}
\frac{\Delta\rho}{\rho}\sim 2\frac{\Delta t}{t}=3H\Delta t\sim\frac{3}{2}\frac{H}{M},
\end{equation}
where $\Delta\rho\equiv |\rho_{f}-\rho_{i}|$. We now focus on the last two wells, such that $\rho \sim \Lambda^4$. Using Friedmann's equation, $3H^2 M_{p}^2 = \rho$, and $M^2 \approx \frac{\Lambda^4}{f^2}$ we find
\begin{equation}
\label{eq:enloss}
\Delta\rho\sim \rho\cdot \frac{\rho^{1/2}}{M_{p} M}\sim \Lambda^4\frac{f}{M_{p}}=\kappa\frac{m^2 f^3}{M_{p}}.
\end{equation} 
We can now quantitatively discuss the probability of having a phase decomposition. To this end we have to compare the decrease in energy density due to friction $\Delta\rho$ and the uncertainty due to field fluctuations $\delta\rho$. If we have $\delta\rho\sim \Delta\rho$, the field will populate more than one vacuum with $O(1)$ probability, as should be clear from Fig.~\ref{fig:energybands}. The term probability here refers to the exact choice of model parameters which, at that level of precision, appears arbitrary to the low-energy effective field theorist.

The task of the next section is therefore to present two possible origins of the fluctuations $\delta\rho$. These are respectively classical inflationary inhomogeneities and quantum uncertainties of the scalar field $\phi$.

\end{subsection}

\end{section}

\begin{section}{Fluctuations and phase decomposition}
\label{sec:phases}

In this section we will analyse two sources of fluctuations of the inflaton field. In Sec.~\ref{sub:evolution} we focus on (classical) inflationary fluctuations. Those originate as sub-horizon size quantum fluctuations, but are then stretched to super-horizon size and become classical. After inflation, they re-enter the horizon and may lead to the phase decomposition described above. In Sec.~\ref{sub:quantumflucts} we focus instead on the intrinsic quantum uncertainty which characterizes a quantum field at any given time. Independently of any inflationary pre-history of our field, this effect is present directly during the oscillatory stage and may also lead to phase decomposition.
In fact, the estimates that we obtain in this section imply a small probability of phase decomposition. However, in Sec.~\ref{sub:numerics} we will see that the probability can be much larger because fluctuations may be enhanced after inflation.

\begin{subsection}{Inflationary fluctuations}
\label{sub:evolution}

The evolution of inflationary fluctuations on sub- and super-horizon scales is well-known (see e.g. \cite{Baumann, Mukhanov}). For us the only crucial point is that, once a certain inflationary mode re-enters the horizon it behaves like a dark matter fluctuation, i.e.~it is a decaying oscillation (see \cite{Ratra:1990me, Hwang:1996xd} for a detailed study of scalar field fluctuations after inflation). From now on we focus on the amplitude of such an oscillation, which we denote by $\delta\phi^{inf}_{k}$. The background is denoted by $\phi_0$ and satisfies the equation of motion \eqref{eq:eom}.

The initial conditions on $\delta\phi_{k}^{inf}$ are determined by matching with the power spectrum of the gauge-invariant curvature perturbation $\mathcal{R}$. This quantity is conserved on superhorizon scales and is given by
\begin{equation}
\label{eq:curvaturespectrum}
\Delta_{\mathcal{R}}^2(k)=\frac{1}{8\pi^2}\left[\frac{1}{\epsilon}\frac{H^2}{M_{p}^2}\right]_{exit} \ ,
\end{equation}
where the right hand side is evaluated at horizon exit, i.e. for $k\sim H$. In the slow-roll regime the quantity $\epsilon$ in \eqref{eq:curvaturespectrum} coincides with the familiar expression $\epsilon=\frac{1}{2}(V'/V)^2$. 

To determine the probability of phase decomposition we need to understand how field fluctuations $\delta \phi_k^{inf}$, once they re-enter the horizon, give rise to density fluctuations $\delta \rho_k^{inf}$. More specifically, we wish to determine $\delta \rho_k^{inf}$ at time $t_{\Lambda}$. This is the time at which the amplitude of the background $\phi_{0}$ has decreased to values comparable to the width of the last wells, and it is given by 
\begin{equation}
t_{\Lambda}\sim H_{\Lambda}^{-1}\sim \frac{1}{\kappa^{1/2}(f/M_{p})m} \ .
\end{equation}
Let us now consider a mode with $k \sim m$. This mode will re-enter the horizon at a time $t_0 \sim m^{-1} < t_{\Lambda}$. At $t_0$ we can determine the size of $\delta \phi_k^{inf}$ by matching with the curvature perturbation. However, the field fluctuation $\delta \phi_k^{inf}$ thus obtained will be out of phase with the oscillation of the background $\phi_0$. It is maximal when $\dot{\phi}_0$ is maximal and vanishes when $\phi_0$ is at a turning point. If this remained the case for the subsequent evolution until $t_{\Lambda}$, the fluctuation could not give rise to a phase decomposition. This can only occur if we have a sizable fluctuation at a turning point of $\phi_0$.

However, we expect decoherence between $\delta \phi_k^{inf}$ and $\phi_0$ after only a few oscillations. The reason is that $\phi_0$ oscillates with frequency $m$ while a mode with $k \sim m$ will oscillate with frequency $\sqrt{k^2 + m^2} \sim \sqrt{2} m$. Thus, at some time after $t_0$ the field fluctuation $\delta \phi_k^{inf}$ will be an admixture of out-of-phase but also in-phase-oscillations w.r.t.~to $\phi_0$. It is exactly the in-phase-oscillations which do not vanish at turning points and it is these fluctuations which give rise to density perturbations $\delta \rho_k^{inf}$. In the following we will thus assume that once a mode enters the horizon, while initially out of phase with $\phi_0$, it will give an $\mathcal{O}(1)$ contribution to an in-phase oscillation with corresponding $\delta \rho_k^{inf}$ after only a few periods.\footnote{As we cannot quantify this effect exactly, we will now drop exact numerical prefactors in all following expressions.}

To estimate the size of fluctuation at time $t_{\Lambda}$ given a fluctuation at $t_0$ we need to take the expansion of the universe into account. In particular, the energy density scales as 
\begin{equation}
\rho \sim a^{-3} \ , \qquad \frac{\delta \rho}{\rho} \sim a \quad \Rightarrow \quad \delta \rho \sim a^{-2} \ .
\end{equation}
Thus, for a density fluctuation with $k \sim m$ there is a dilution in the time span between $t_0$ and $t_{\Lambda}$. 

Let us now determine the probability of phase decomposition due to a mode with $k \sim m$. As argued before, the field fluctuation will quickly give rise to a density fluctuation. Instead of taking the intermediate step via field fluctuations, let us match the density fluctuations directly to the curvature fluctuations at re-entry:
\begin{equation}
\Delta_{\mathcal{R}}^2 \sim \frac{\Delta^{2}_{\delta\rho}}{\rho^2} \quad \Rightarrow \quad \delta \rho_k^{inf}(t_0) \sim \rho(t_0) \sqrt{\Delta_{\mathcal{R}}^2} \sim \frac{\rho(t_0)}{M_p} \left[\frac{H}{\epsilon^{1/2}}\right]_{exit} \ .
\end{equation}
Now, using 
\begin{equation}
\rho(t_0) \sim m^2 M_p^2 \ , \qquad \left[\frac{H}{\epsilon^{1/2}}\right]_{exit} \sim m \ , \qquad \frac{a^{-2}(t_{\Lambda})}{a^{-2}(t_0)} \sim \kappa^{2/3} (f/M_p)^{4/3}
\end{equation}
we obtain
\begin{equation}
\label{eq:probinf}
\frac{\delta\rho^{inf}}{\Delta\rho}\sim \kappa^{-1/3} \left(\frac{m}{M_{p}}\right) \left(\frac{M_{p}}{f}\right)^{5/3}. 
\end{equation}
The above probability was derived for modes with $k \sim m$. However, we are interested in the situations when the above probability is largest. Thus let us consider how this result is modified if we consider a mode that re-enters the horizon earlier or later. Modes with $k>m$ enter the horizon earlier and will hence experience more dilution. As a result, the corresponding probability of phase decomposition will be smaller. Modes with $k<m$ enter the horizon later and will be diluted less. Hence they would in principle give rise to a larger probability than \eqref{eq:probinf}. However, they cannot enter too late as they need to have enough time to decohere w.r.t.~$\phi_0$. A more detailed analysis would be needed to determine how late a mode can enter the horizon and nevertheless give rise to a sizable density fluctuation. Thus, \eqref{eq:probinf} should be seen as a reasonable estimate. We shall comment more on the size of this probability at the end of Sec.~\ref{sub:quantumflucts}.
\end{subsection}

\begin{subsection}{Quantum fluctuations}
\label{sub:quantumflucts}

There is another potentially relevant source of fluctuations of the field $\phi$. This is the intrinsic uncertainty due to the quantum nature of our scalar. It can be simply written as
\begin{equation}
\label{eq:uncert}
\delta\phi^{q}_{k}\sim k.
\end{equation}
In order to estimate the maximal effect of these fluctuations, let us consider the following setting: consider a scalar field $\phi$ with fluctuations $\delta\phi^{q}_{k}$ approaching the local maximum of some potential. This is basically as in Fig.~(\ref{fig:dw2}), with the field approaching the maximum from the left side. We then ask the following question: what is the field distance from the local maximum which the background field has to reach such that its fluctuations can lift it over the potential barrier? The relevant energy scale around the maximum is $O(M)$, where $M^2$ is the curvature of the potential at the maximum. One can convince oneself that only fluctuations of the order $\delta\phi\sim M$ are relevant for overcoming the barrier: modes with $k \gg M$ have an effective non-tachyonic mass and are insensitive to the instability. In contrast, modes with $k\ll M$ are sensitive to the tachyonic instability, but their fluctuations are smaller than those of modes with $k\sim M$.

Alternatively, we can understand this point by considering tunnelling. Hence we take a homogeneous scalar field approaching the maximum and study the conditions under which quantum tunnelling to the other side of the barrier becomes efficient. In order to answer this question, let us use the following standard tunnelling formulae: the tunnelling rate is given by $e^{-S_{0}}$, where $S_{0}$ is the action of critical bubble formation. When the thin wall approximation is applicable, this reads (see e.g. \cite{Coleman})
\begin{equation}
\label{eq:criticalaction}
S_{0}=\frac{27\pi^{2}\sigma^{4}}{2(\Delta V)^{3}}=\frac{27\pi^2\delta\phi^{2}}{2 M^2},
\end{equation}
where $\sigma\sim M\delta\phi^2$ is the bubble wall tension and $\Delta V$ can be estimated with a quadratic approximation, $\Delta V\sim M^2\delta\phi^2$. However, in most of the cases the thin-wall calculation is not appropriate. Nevertheless, we still expect $S_{0}\sim \delta\phi^2/M^2$, with a different prefactor.\footnote{We have checked the behaviour for an inverted parabola. In the vicinity of the maximum this is  a reasonable approximation, with a prefactor of the order of $10^{4}$.} 
According to \eqref{eq:criticalaction} the tunnelling rate is unsuppressed when $\delta\phi\gtrsim 10^{-1} M$, which is the same condition we found with the previous approach up to a numerical prefactor. Given the uncertainty in the prefactor, for the time being we use the parametric dependence $\delta\phi\sim M$.

Following these two arguments the uncertainty in the energy density induced by such field fluctuations is
\begin{equation}
\label{eq:quantumgain}
\delta\rho^{q}\sim M^2\delta\phi^{q 2}_{k}\sim M^4.
\end{equation}
If quantum fluctuations induce an energy gain which is larger than the loss due friction one expects phase decomposition.
Therefore, we compare \eqref{eq:quantumgain} with \eqref{eq:enloss}, using also $M^2\simeq \Lambda^4/f^2$. We obtain
\begin{equation}
\label{eq:probquant}
\frac{\delta\rho^{q}}{\Delta\rho}\sim \kappa\left(\frac{m}{M_{p}}\right)^2\left(\frac{M_{p}}{f}\right)^3.
\end{equation}
Comparing \eqref{eq:probinf} and \eqref{eq:probquant} we conclude that the probability of a phase decomposition due to inflationary fluctuations is larger than the one due to quantum ones by at least a factor $(f/M_{p}\kappa)^{4/3}(M_{p}/m)$. 

Let us now discuss the size of the probabilities \eqref{eq:probinf}, \eqref{eq:probquant}. In quadratic inflation, $m\sim 10^{-5}M_{p}$. For $\kappa\sim O(10)$ and $f \sim 10^{-13/5} M_p$ the probability \eqref{eq:probinf} is of the order of $0.1$. This implies that a phase decomposition is rather likely for these values of parameters. For the same choices, the probability \eqref{eq:probquant} is only slightly smaller, i.e. $\mathcal{P}^{q}\sim 10^{-6/5}$. However, further numerical suppression is expected in \eqref{eq:probquant}. Therefore we conclude that in the regime $\kappa\sim O(10)$, $f\lesssim 0.3\cdot 10^{-2} M_{p}$ phase decomposition is likely to happen as a consequence of inflationary fluctuations. If $f$ grows above $10^{-2}$ phase decomposition quickly becomes improbable. Whether such small values of $f$ are natural depends on the details of the model leading to \eqref{eq:potential}. We will comment more on the size of $f$ at the end of Sec.~\ref{sub:numerics}.

Further there exist arguments \cite{Hebecker:2015zss} based on the Weak Gravity Conjecture (WGC) \cite{ArkaniHamed:2006dz} which constrain the size of modulations of the potential in axion monodromy inflation (for an earlier somewhat different perspective see \cite{Ibanez:2015fcv}). It is thus important to check whether the region of parameter space considered in this work is consistent with these bounds. Given a domain wall with tension $T$ and charge $e$ the electric WGC demands $T \lesssim e M_p$. In our case we have $T=\Lambda^2 f $ and $e= 2 \pi m f $ (see \cite{Hebecker:2015zss} for details). Using our definition $\Lambda^2 = \sqrt{\kappa} mf$ the WGC bound reads
\begin{equation}
\sqrt{\kappa} \lesssim \frac{M_p}{f} \ .
\end{equation}
As a result, while $\kappa$ is bounded from above, our preferred parameter region for phase decomposition ($\kappa \sim \mathcal{O}(10)$, $f \lesssim 10^{-2} M_p$) is consistent with the WGC. Interestingly, let us notice that the region of parameter space which is ruled out by the WGC is also constrained by current observations, which require $\Lambda^4/(m^2\phi ^2)\lesssim (10^{-3}-10^{-2})$ during inflation\cite{Peiris:2013opa, Meerburg:2013dla}, where $\phi \gtrsim M_{p}$.

To close this section let us make two important remarks. First, note that phase decomposition can still occur even if the probability is small. In this case we only expect very few bubbles per Hubble patch. The second comment concerns once again the distinction between classical and quantum fluctuations. These two sources can also be distinguished based on the length scale $R$ at which their effect is strongest. As we explained, classical inhomogeneities are most relevant at $R\sim H_{\Lambda}^{-1} \sim M_p/ \Lambda^2$ while the quantum effect is strongest for $R\sim M^{-1} \sim f/ \Lambda^2$. Since $f<M_{p}$, the size $R$ of the latter inhomogeneities is parametrically smaller than that of the inflationary ones. 
\end{subsection}

\end{section}

\begin{section}{Enhancement of fluctuations}
\label{sub:numerics}

Until now we have assumed that fluctuations, whether they are of classical or quantum nature, remain small during the evolution of the universe after inflation. The aim of this section is to discuss a possible enhancement of the fluctuations $\delta\phi_{k}$ due to the functional form of the potential \eqref{eq:potential}. Before going into details, let us summarise the main result. In this section we are mainly interested in fluctuations with $k \sim m$ at time $t_{\Lambda}$. When the field oscillates at the bottom of the potential containing only the last few well, these fluctuations can be enhanced for certain values of $f$ and $\kappa$. Crucially, these modes never exit the horizon during inflation, because at $t<t_{\Lambda}$ their wavelength is smaller than the Hubble radius. Therefore, these modes are never classicalised. Nevertheless, in this section we study their enhancement treating them as classical. We expect that our analysis will still capture the main effect. We provide a more detailed discussion at the end of this section.

A large growth of fluctuations can severely affect our conclusions. On the one hand, large fluctuations of the inflaton field may be desirable to some degree in our setup: the larger $\delta\phi_{k}$, the easier it is to cross the barrier between two local minima. Furthermore, if a mode with $k\gg m$ has a large amplitude, it may induce a phase decomposition independently of the modes with $k\sim m$ that we have studied in the previous section. If this is the case \eqref{eq:probinf} underestimates the probability for phase decomposition.

On the other hand, large fluctuations with $k\gg m$ can lead to short-range violent dynamics rather than to the formation of well defined bubbles (which need a length scale $\gtrsim 1/M$). As we will describe in more detail in Sec.~\ref{sec:gravitationalwaves}, this can negatively affect the strength of the gravitational wave signal related to our setting. 

For these reasons, it is crucial to assess if any large growth of fluctuations occurs in our setting. 	As we have already mentioned, we focus on the enhancement of classical fluctuations and comment later on the applicability of the results for quantum modes.
Such an analysis involves solving the coupled equations of motion for the background field and for the fluctuation in an expanding background and in the presence of a non-zero gravitational field. In Appendix~\ref{sub:nonlinearities} we present a first step towards this goal by providing the relevant equations of motion.

Here we will perform a somewhat different, simplified analysis, assuming that the gravitational field is negligible for the following reason. As we describe in Appendix \ref{sec:after}, the addition of a gravitational field leads to a dark matter-like growth of fluctuations, which is negligible compared to the exponential growth that we
are seeking in this section. Let us rewrite the equations of motion in a way that is more suitable for a numerical analysis. Namely, we define $t'=mt$ and $\varphi_{0}=\phi_{0}/f$. Then the linearised equations of motion without gravity read: 
\begin{align}
\label{eq:backgroundnum}
\varphi''_{0}+\frac{2}{t'}\varphi'_{0}+\varphi_{0}-\kappa\sin\left(\varphi_{0}\left(t' \right)\right)&=0\\
\label{eq:fluctuation2}
\delta\phi''_{k}+\frac{2}{t^{'}}\delta\phi'_{k}+\left[1+\frac{k^{2}}{m^{2}a^{2}(t')}-\kappa\cos\left(\varphi_{0}\left(t'\right)\right)\right]\delta\phi_{k}\left(t'\right)&=0,
\end{align}
where `prime' now denotes a derivative w.r.t.~$t'$ and where we have used $\frac{a'}{a}=\frac{2}{3t'}$ during matter domination. 

In the absence of friction, the background solution $\varphi_0$ is periodic and the equation of motion for $\delta \phi_k$ is a Hill's equation. Solutions to such an equation exhibit a resonant behaviour for certain values of $k$ \cite{Whittaker,Jaeckel:2016qjp}. This is similar to the resonances encountered in the context of preheating (for a review see \cite{Kofman:1997yn}). However, the inclusion of a time-dependent background as well as friction, may affect the growth of the solution. 
Generically, one expects that modes with $k\gtrsim m$ may experience exponential growth for certain values of $f$ and $\kappa$. We investigated the behaviour of $\phi_0$ and $\delta\phi$ numerically, for certain values of the parameters.
In Fig.~\ref{fig:background100} the background $\phi_0$ is plotted as a function of $t$ for $f=10^{-2}M_{p}, \kappa=60$. For this parameter choice the field is caught near one of the local minima after only one oscillation. 

\begin{figure}
    \centering
    \includegraphics[width=0.5\textwidth]{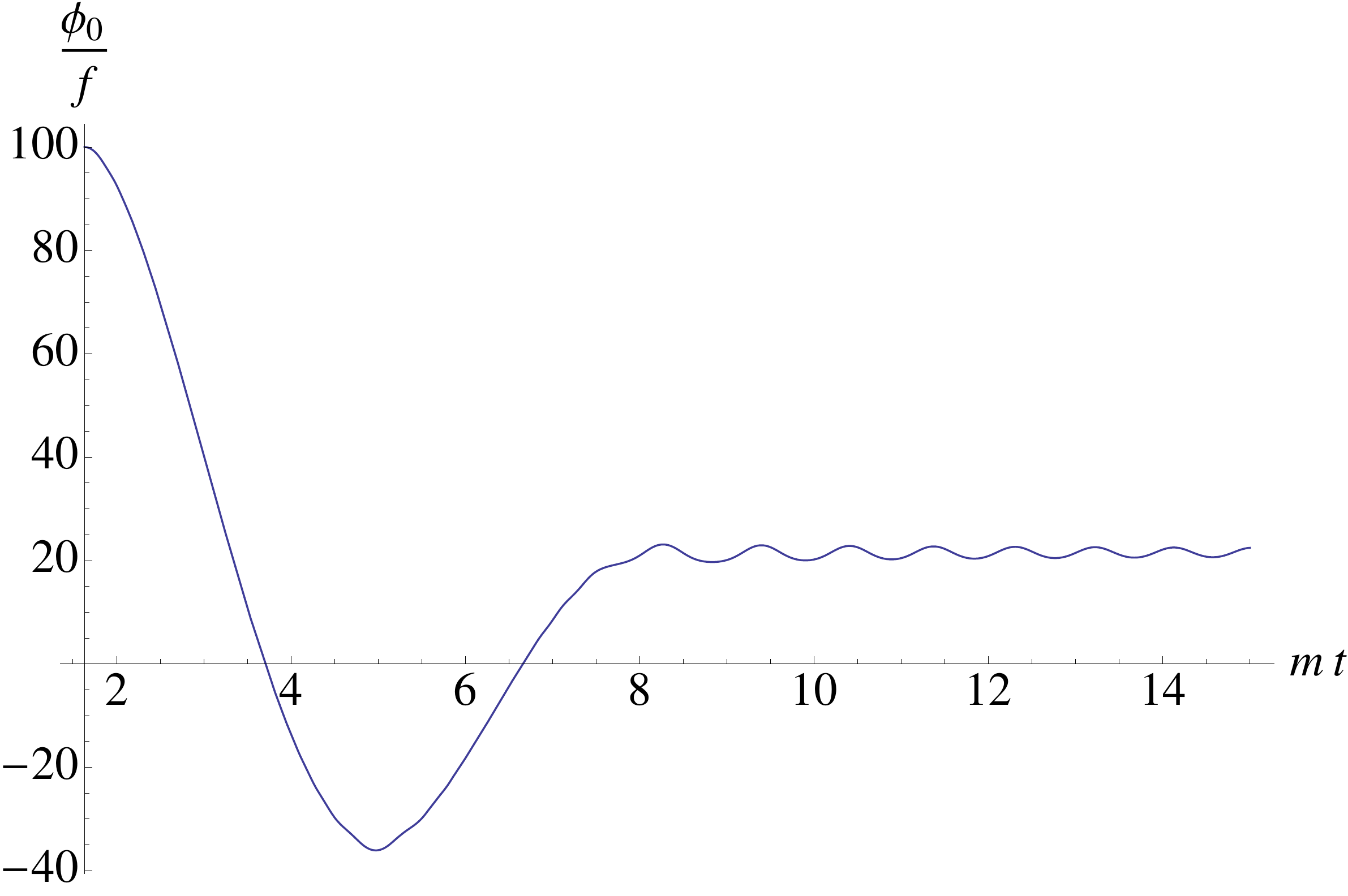}
    \caption{Evolution of the background $\phi_{0}\left(t'\right)$, according to \eqref{eq:backgroundnum}. The initial conditions $\phi_0\left(t'_{i}\right)=M_{p}$ and $t_{i}=2\sqrt{2}/(\sqrt{3}m)$ are determined by violation of the slow roll condition. Furthermore  $\phi'_0\left(t'_{i}\right)=0$, Here $f=10^{-2}M_{p}, \kappa=60$. The field is caught in one of the cosine wells around $t=8/m$.}
    \label{fig:background100}
\end{figure}

For the same values of $f$ and $\kappa$ we plot the evolution of the mode $\delta\phi_{5m}$ up to the time when the field is caught in a local minimum in Fig.~\ref{fig:fluctuation100}. We see that before the field settles in one of the local minima, this mode does not grow. Afterwards, the field is oscillating in an approximately quadratic potential and therefore we do not expect any growth. Note that our equations are homogeneous in $\delta\phi_{k}$. Hence our numerical determination of the enhancement is not affected by the initial value of $\delta\phi_{k}$.

\begin{figure}
    \centering
    \includegraphics[width=0.5\textwidth]{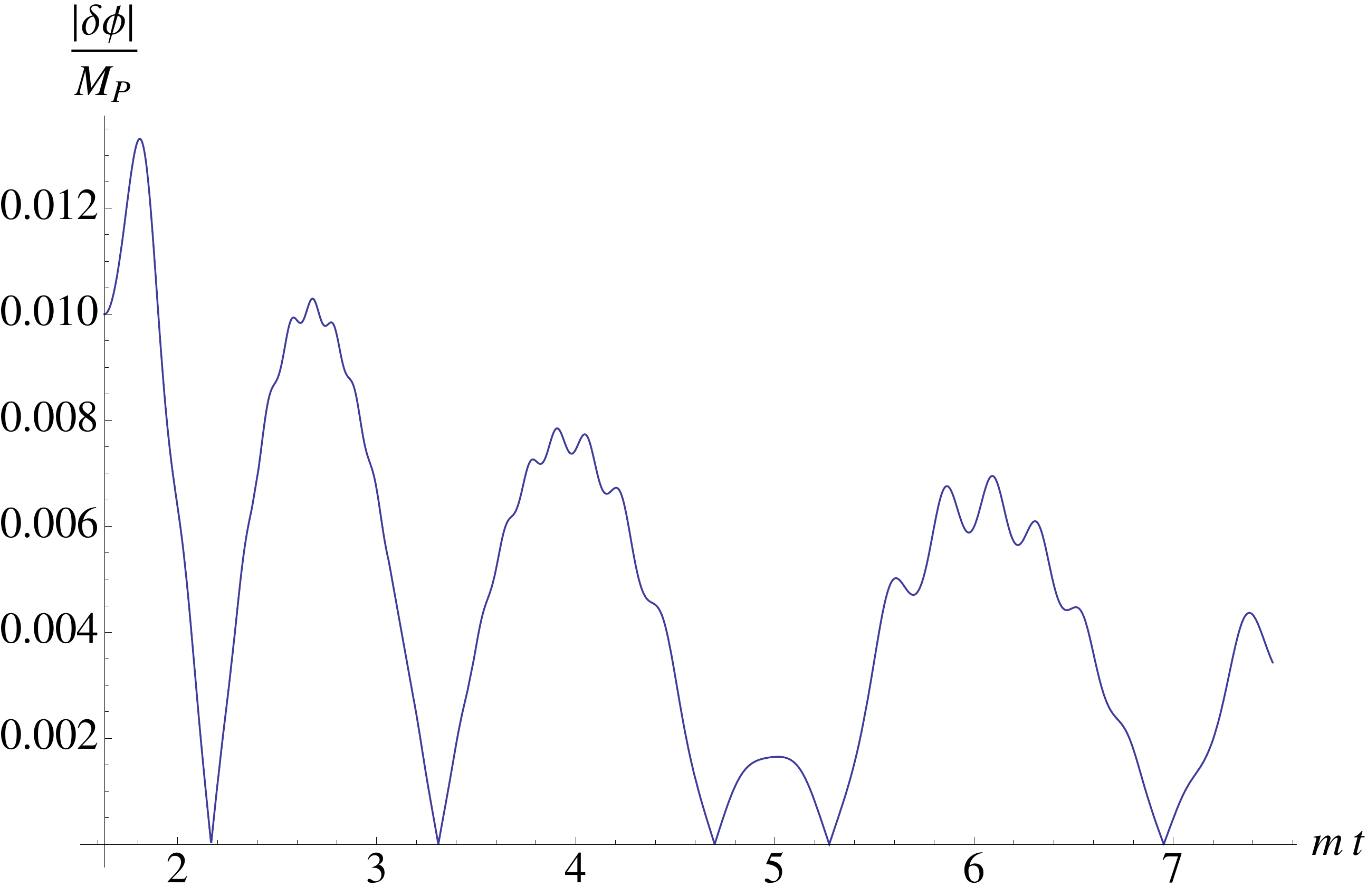}
    \caption{Evolution of the fluctuations $\delta\phi_k\left(t'\right)$, according to \eqref{eq:fluctuation2}. We have chosen: $k=5m, \delta\phi_k '\left(t'_{i}\right)=0$, and $t_{i}=2\sqrt{2}/(\sqrt{3}m)$. Furthermore $ f=10^{-2} M_{p}, \kappa=60$. After $t\approx 8/m$, $\phi_0$ is stuck in one of the cosine wells.}
    \label{fig:fluctuation100}
\end{figure}

However, the situation can be radically different, as the next numerical example shows. In Fig.~(\ref{fig:background300}) we plot the background scalar field for $f=M_{p}/300, \kappa=20$. The field gets stuck in one of the cosine wells after six oscillations. 
\begin{figure}
    \centering
    \includegraphics[width=0.5\textwidth]{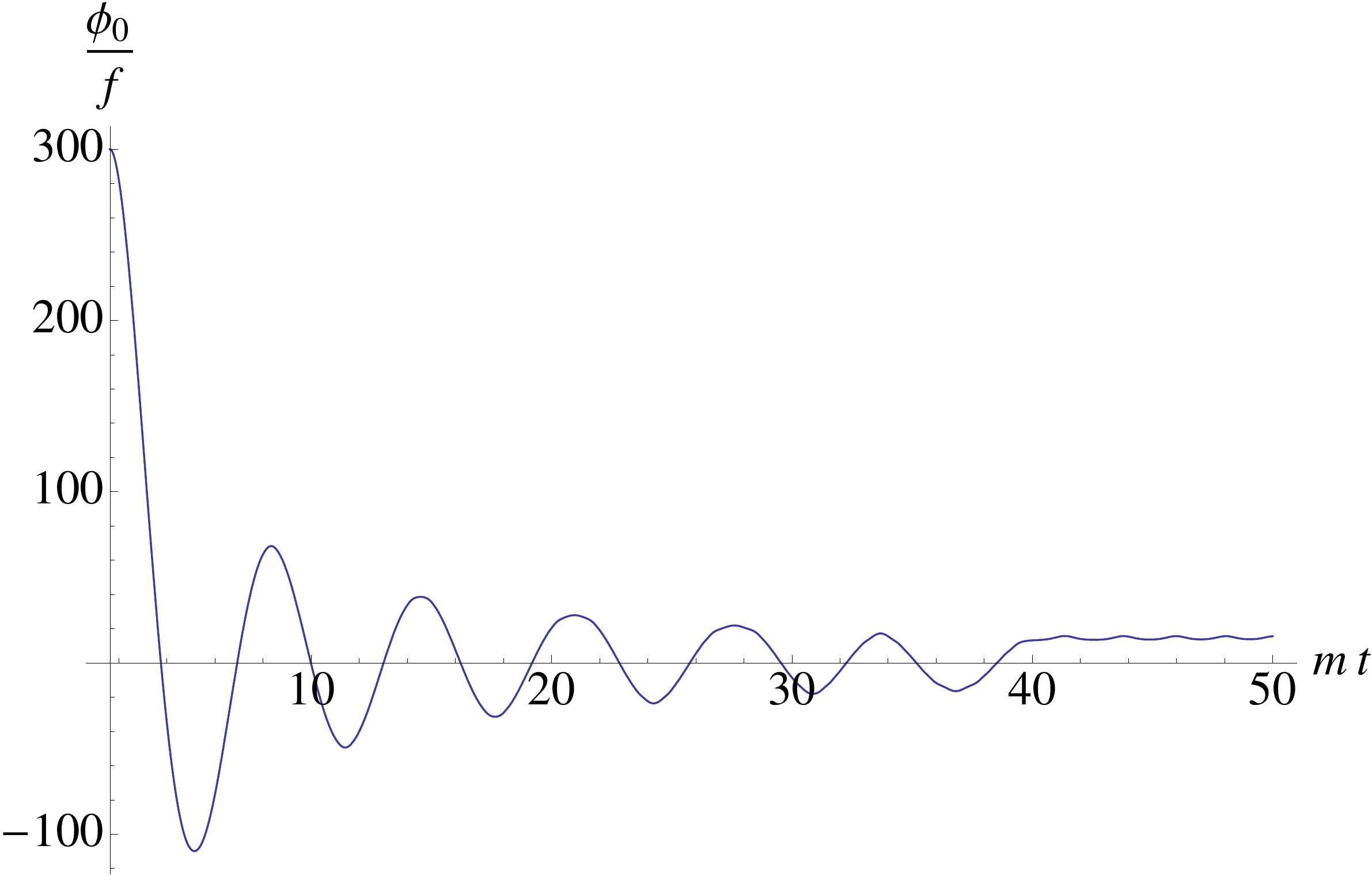}
    \caption{Evolution of the background $\phi_{0}\left(t'\right)$, according to \eqref{eq:backgroundnum}. The initial conditions $\phi_0\left(t'_{i}\right)=M_{p}$ and $t_{i}=2\sqrt{2}/(\sqrt{3}m)$ are determined by violation of the slow roll condition. Furthermore $\phi'_0\left(t'_{i}\right)=0$. Here $f=M_{p}/300, \kappa=20$. The field is caught in one of the cosine wells around $t=40/m$.}
    \label{fig:background300}
\end{figure}
As a consequence of the longer time that the field spends oscillating across several cosine wells, fluctuations can now grow significantly . In Fig.~(\ref{fig:fluctuation300}), we plot the logarithm of the absolute value of $\delta\phi_{k}$, again for $k=5m, f=M_{p}/300, \kappa=20$. We see that the amplitude of this mode grows by three orders of magnitude before the background field settles in one of the local minima. In fact, such strong growth takes us out of the regime of validity of the linearized equation of motion (as $\delta \phi_k$ becomes larger than $f$).

\begin{figure}
    \centering
    \includegraphics[width=0.5\textwidth]{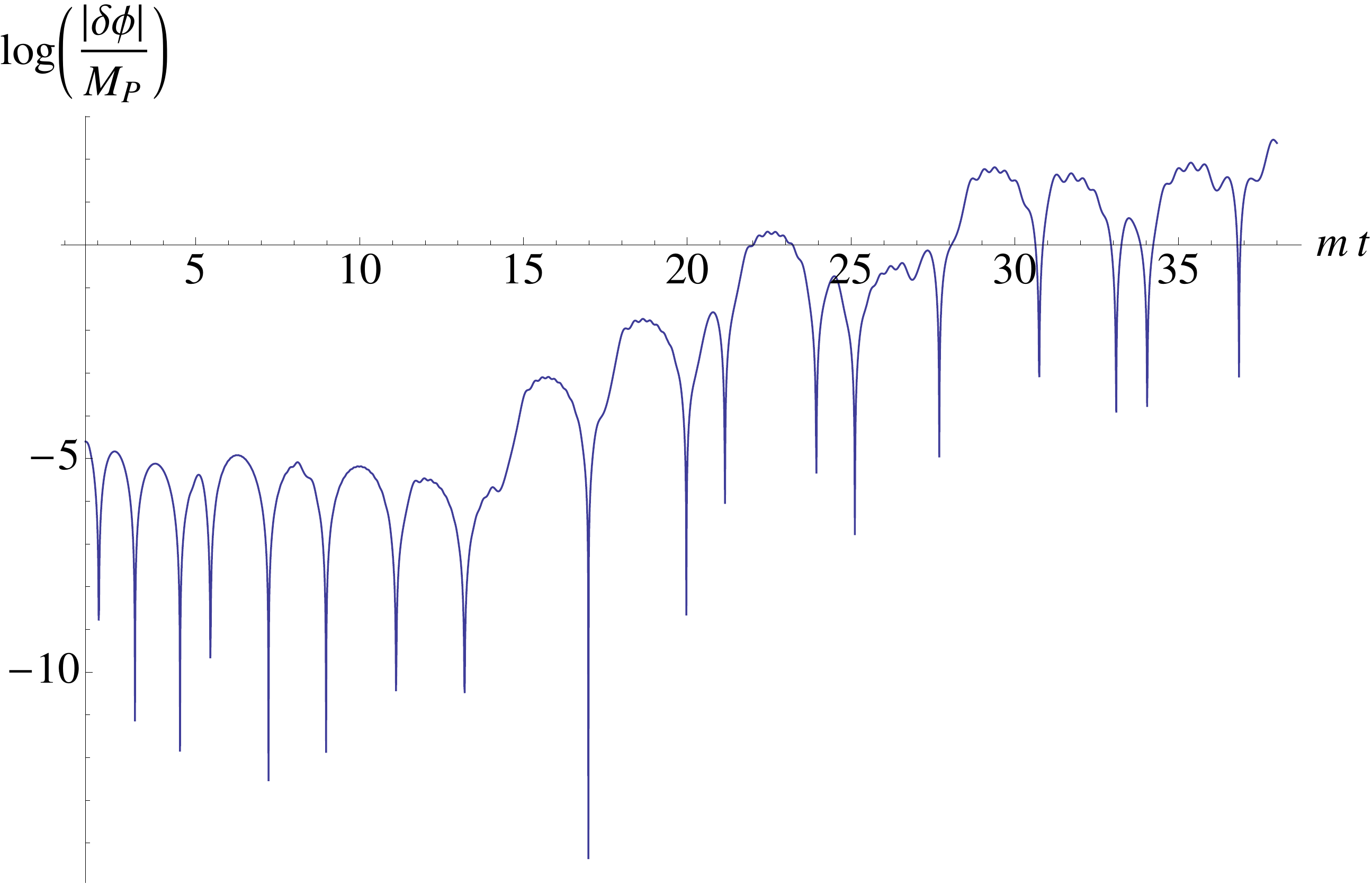}
    \caption{Logarithmic evolution of the absolute value of $\delta\phi_k\left(t'\right)$, according to \eqref{eq:fluctuation2}. We have chosen: $k=5m$, $\delta\phi_{k}'\left(t_{i}\right)=0$, and $t_{i}=2\sqrt{2}/(\sqrt{3}m)$. Furthermore $f=M_{p}/300, \kappa=20$. After $t\approx 40/m$, $\phi_0$ is stuck in one of the cosine wells.}
    \label{fig:fluctuation300}
\end{figure}

Recently, the growth of fluctuations in a potential with cosine modulations was studied in \cite{Jaeckel:2016qjp}. The authors argue that, for a relatively small Hubble scale and neglecting gravity, the fluctuations for $k\sim m$ can grow as $e^{N_{k}}$, where $N_{k}$ is roughly given by
\begin{equation}
N_{k}\sim \frac{m}{H_{k\sim m}}F(\kappa).
\end{equation}
Here $F(\kappa)$ is a function of the order up to a few whose value depends on the initial amplitude of $\phi$.
In our case, since $H_{k\sim m}\sim \Lambda^{2}/M_{p}\sim \kappa^{1/2} m (f/M_{p})$, we conclude that fluctuations should grow with exponent:
\begin{equation}
\label{eq:exponent}
N_{k\gtrsim m}\sim \frac{M_{p}}{f},
\end{equation}
where we have dropped any $\kappa$-dependence due to our ignorance regarding $F(\kappa)$.
Our numerical examples, which take Hubble expansion into account explicitly, confirm that for small $f$ such an exponential growth does happen for most values of $\kappa\gtrsim 10$. In contrast, we do not observe enhancement if the value of $f$ is chosen too large.
 
Apart from this qualitative discussion, we are unfortunately unable to provide an analytical understanding of the dependence of the enhancement on the parameters $\kappa$ and $f$. This is partly due to the fact that the phenomenon strongly depends on the precise minimum the field ultimately settles in. The latter question can only be addressed in a probabilistic approach, i.e. we can only say where the field is more \emph{likely} to get trapped. Therefore, we do not have a precisely monotonic dependence of the enhancement in terms of $f$ and $\kappa$.

In the absence of an analytical treatment, we performed a numerical search for enhanced fluctuations, focusing on modes with $k\gtrsim m$ at the time when $H\sim O(\Lambda^{2}/M_{p})$. The results are reported in Fig.~\ref{fig:resonance} in the form of a grid of points. Each point corresponds to a value of $\kappa$ and $f$. We observe that in the region of interest fluctuations tend to be enhanced whenever $f\lesssim M_{p}/200$. Here, we define `enhancement' as follows: we will refer to a mode to be enhanced if its original amplitude has grown by roughly two orders of magnitude before getting caught. The enhanced $\delta\phi_{k}$ is comparable to $f$ and we are therefore at the boundary between the linear and non-linear regime. Interestingly, this boundary corresponds to parameter values such that the probability \eqref{eq:probinf} is of the order $10^{-1}$. However, note that the probability in \eqref{eq:probinf} was determined for modes which will exhibit $k < m$ at $t_{\Lambda}$. While such fluctuations may experience growth, the generic expectation is that enhancement does not occur for modes with $k <m$. 

\begin{figure}
\centering
\includegraphics[width=0.5\textwidth]{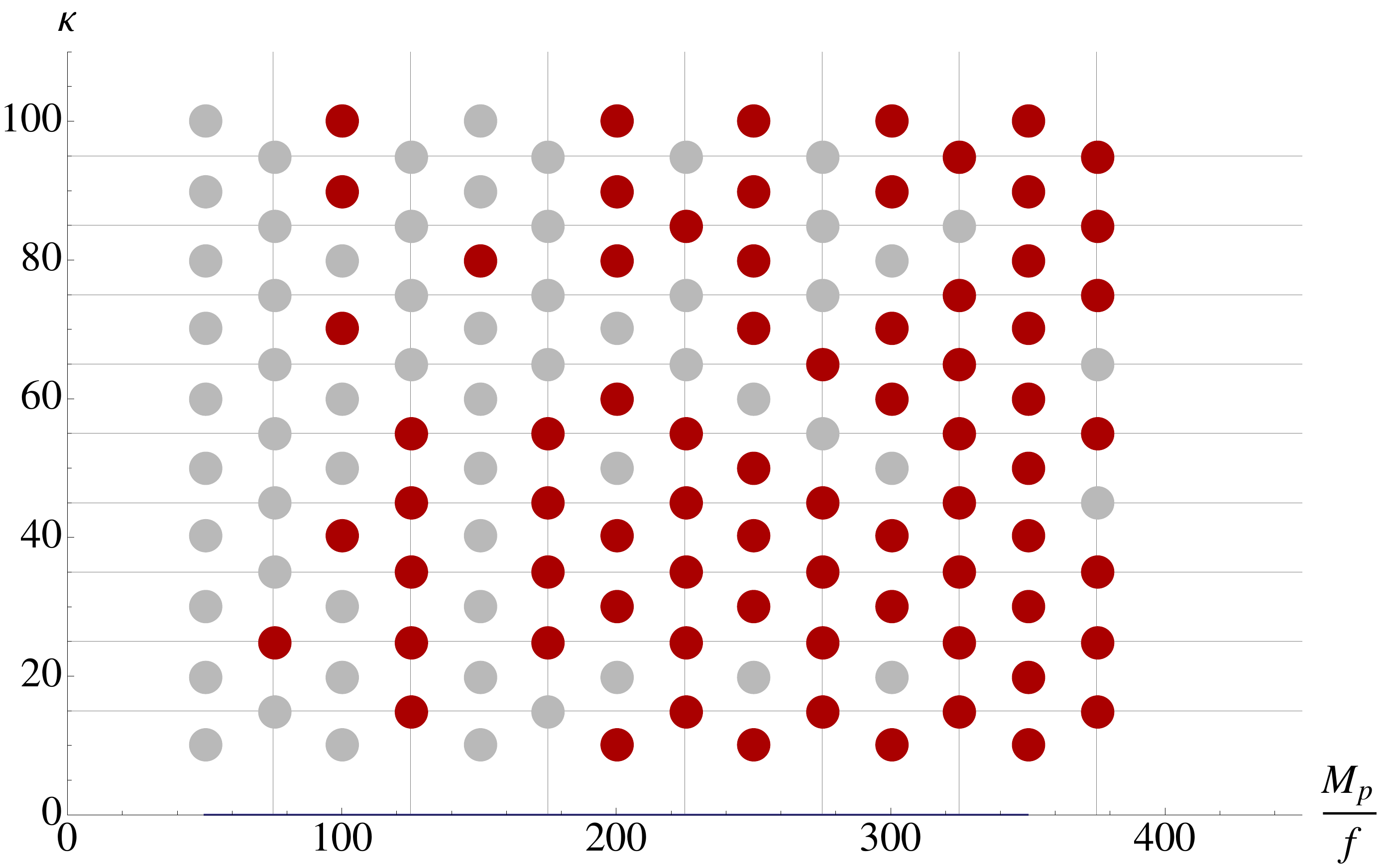}
\caption{Grid showing for which values of $\kappa$ and $f$ resonance occurs. Grey points correspond to the case of no (or too small) enhancement. Red points correspond to large enhancement. By the latter we mean that fluctuations grow by at least two orders of magnitude before getting caught near a local minimum. The equations of motion are solved for $k= M=\kappa^{1/2}m$ at the time when $H\sim O(\Lambda^2/M_{p})$.}
    \label{fig:resonance}
\end{figure}
Let us now briefly discuss enhancement of quantum fluctuations. This is in principle a complicated issue: we cannot use the classical equations of motions to analyse the behaviour of the quantum system. However, it is important to notice that, if quantum fluctuations are initially enhanced, they quickly become classical. Here by ``classical'' we mean that their occupation number becomes large, such that \eqref{eq:fluctuation2} can be used to study their evolution. Parametric resonance in quantum mechanics and quantum field theory has been studied analytically and numerically: the conclusion is that quantum modes do experience exponential growth (see e.g.~\cite{Weigert, Berges:2002cz}). We expect that the same effect will occur also in the system analysed here. Therefore, our study of classical fluctuations should extend, at least partially, to quantum fluctuations. If more enhancement occurs in the quantum case, then phase decomposition is even more likely. We leave a more detailed study of this effect for future work.

Finally, let us summarise our findings concerning the probability of phase decomposition:
\begin{enumerate}
\item{For $f\gtrsim 0.5\cdot 10^{-2}M_{p}$ enhancement is generically not observed. The probabilities \eqref{eq:probinf} and \eqref{eq:probquant} are small, so that a phase decomposition is unlikely. Nevertheless, it is still possible that very few bubbles of tiny size containing the state of lower energy are nucleated. As we describe in the next section, observational signatures from such a situation may be quite strong. Classical inflationary fluctuations are the dominant cause of phase decomposition in this regime.}
\item{For $f\lesssim 0.5\cdot 10^{-2}M_{p}$ we have the following situation. On the one hand, fluctuations with $k\sim m$ at $t=t_{\Lambda}$ are generically enhanced. These modes are genuinely quantum modes, since they never exited the horizon. In this region, the enhancement may be just large enough to give rise to a probability of phase decomposition of order $O(1)$. Furthermore, we observe numerically that, at fixed $f$ and $\kappa$, modes with $k\gg m$ do not experience the same exponential growth. This will turn out to be a useful observation when examining the gravitational wave signal from the associated phase transition.  

On the other hand, according to \eqref{eq:probinf} and \eqref{eq:probquant}, classical and quantum modes with $k \lesssim m$ can lead to a phase decomposition, even if they are not enhanced. Therefore we conclude that a phase decomposition is very likely to be induced. Assuming that our analysis of enhanced classical fluctuations extends to the quantum ones, the dominant cause of phase decomposition are quantum modes with $k\sim m$ at $t_{\Lambda}$.}
\item{For $f\ll 10^{-2} M_{p}$ fluctuations with $k\sim m$ are strongly enhanced. In this region phase decomposition is very likely to occur. However, it is hard to provide any description of such a highly non-linear regime. Classical and quantum fluctuations with $k\lesssim m$ are generically not enhanced, but would also lead to phase decomposition according to \eqref{eq:probinf} and \eqref{eq:probquant}.}

\end{enumerate}
One more comment is in order before moving on to the phenomenological signatures of our setup. Phase decomposition happens generically for rather small axion decay constants. One may question whether such values of $f$ are plausible in the spirit of axion monodromy. The answer depends very much on the framework in which monodromy is implemented. In a stringy setup it is a question of moduli stabilisation: e.g. in the Large Volume Scenario (LVS) \cite{Balasubramanian:2005zx, Conlon:2005ki} decay constants are generically suppressed by the volume of the compactification manifold and are therefore naturally small.
\end{section}

\begin{section}{Gravitational radiation from Phase Transitions}
\label{sec:gravitationalwaves}

In the previous sections, we have described a mechanism which can potentially lead to phase decomposition in the early universe after inflation. In this section, we will assume that such phase decomposition indeed occurs. In the presence of different populated vacua, bubbles containing the state of lowest energy can form and expand. Their collisions are a very interesting and well-known source of gravitational radiation. This has been studied in detail in the literature in various contexts and regimes (see \cite{Caprini:2015zlo} and references therein).

The aim of this section is twofold. First of all, we would like to elucidate the peculiarities of our setup concerning the energy released into gravitational waves during the collision of bubbles. Rather than focusing on precise calculations, we will give a qualitative discussion and provide formulae analogous to the more familiar case of bubbles colliding in a relativistic plasma. In this case, there are three possible sources of gravitational radiation: the collision of bubble walls, sound waves in the plasma and its turbulent motion. The second goal of this section is to give estimates of the relic density and frequency of the gravitational wave signal which can be obtained in our setup.

\begin{subsection}{Gravitational waves from bubble collision}

The focus of this subsection is the collision of bubbles and the possible shocks in the fluid surrounding them. These phenomena are usually studied in the so-called \emph{envelope} approximation \cite{Kosowsky:1991ua}. The latter consists in assuming that the energy liberated in gravitational waves resides only in the bubble walls before the collision. Furthermore, it is assumed that only the uncollided region of those walls contributes to the production of gravitational waves, i.e.~the interacting region is neglected. Such an approximation has been initially applied to the case of vacuum-to-vacuum transitions \cite{Kosowsky:1992rz} and later to collisions in a radiation bath.

In a thermal phase transition, the energy released into gravitational waves depends on four parameters. First of all, there is the time scale of the phase transition $\delta^{-1}$ or, equivalently, the initial separation between two bubbles $d\sim \delta^{-1}$. Secondly, there is the ratio $\eta$ of the vacuum energy density $\epsilon$ released in the transition to that of the thermal bath, i.e.
\begin{equation}
\eta\equiv \frac{\epsilon}{\rho_{rad}^{\star}},
\end{equation}
where $\star$ specifies that the quantity is evaluated at the time of completion of the phase transition. Thirdly, the efficiency factor $\lambda$ characterizes the fraction of the energy density $\epsilon$ which is converted into the motion of the colliding walls. Finally, the bubble velocity $v_{b}$ is not necessarily luminal, as the walls have to first displace the fluid around them. The energy released into gravitational waves of peak frequency is then given by \cite{Grojean:2006bp}:
\begin{equation}
\label{eq:grojean}
\frac{\rho_{GW}}{\rho_{tot}}\sim \theta\left(\frac{H_{\star}}{\delta}\right)^2\lambda^2\frac{\eta^2}{(1+\eta)^2}v_{b}^{3},
\end{equation}
where $\rho_{tot}$ is the background energy density at completion of the phase transition. The parameters $v_{b}$ and $\lambda$ are actually expected to be functions of $\eta$, in such a way that for $\eta\sim O(1)$, also $\lambda, v_{b}\sim O(1)$.

In our case, bubbles collide before reheating, therefore there is no radiation bath around them. However, as we describe in Appendix \ref{sec:after}, an oscillating scalar field corresponds to the presence of a matter fluid. Crucially, the time scale of the field oscillations is set by $m^{-1}$, and may be smaller than the time of collision. Therefore, oscillations of the scalar field cannot be generically neglected. Unfortunately, we do not have specific formulae for this case. Since we are interested only in an order of magnitude estimate for the spectrum of gravitational waves, it seems reasonable to extend \eqref{eq:grojean} to our setup, with the obvious modification
\begin{equation}
\label{eq:eta}
\eta\equiv \frac{\epsilon}{\rho_{matter}^{\star}}.
\end{equation}
Furthermore, we shall hide our ignorance about the dependence of $\lambda$ and $v_{b}$ on $\eta$ by defining $\theta_0=\theta\lambda^2v_b^3$ and leave the determination of these parameters for future work. Therefore, we base our estimates on the following formula for the energy released in gravitational waves from the collision of bubbles and shocks in the matter fluid:
\begin{equation}
\label{eq:ours}
\frac{\rho_{GW}}{\rho_{tot}}\approx \theta_{0}\left(\frac{H_{\star}}{\delta}\right)^2\frac{\eta^2}{(1+\eta)^2},
\end{equation}
where in our case $\rho_{tot}\sim \Lambda^4$.
In addition, the peak frequency of gravitational waves in the envelope approximation is given by
\begin{equation}
\label{eq:frequencypeak}
\omega_{peak}\simeq \sigma\delta,
\end{equation}
where $\sigma\lesssim O(0.1)$ should be fixed numerically and includes effects due to subluminal bubble walls velocity.

The next task is to estimate $\delta$ and $\eta$. Let us start with the ratio $H_{\star}/\delta$. The energy density at the time of the phase transition corresponds to the height of the barrier separating the two minima, therefore $H_{\star}\sim \rho^{1/2}/M_{p}\sim \Lambda^2/M_{p}$. We expect the typical frequency of the phase transition to be set by the momentum $k$ of the spatial inhomogeneities of the field $\phi$. The phase transition can be induced by any mode which is present at $t=t_{\Lambda}$. The largest frequency that one can take is set by $k\sim M$, as we have already discussed in Sec. \ref{sub:quantumflucts}. This corresponds to a scenario where phase decomposition is likely. However, bubble collisions are most violent when the field makes it over the barrier separating the two minima only very rarely. In this case there are only few bubbles per Hubble patch. This latter scenario gives the strongest signal as it corresponds roughly to
\begin{equation}
\label{eq:delta}
\frac{H_\star}{\delta}\sim O(1).
\end{equation}
In order to understand how strong can the signal be in our setup, we assume \eqref{eq:delta} in what follows, but one should keep in mind that this is optimistic.

The estimate of $\eta$ is less straightforward, at least conceptually. If we adopt the envelope approximation, then we need to compute the vacuum energy density released in the phase transition. This is simply the difference $\epsilon$ between the energy density of the two minima in Fig.~\ref{fig:two}. Using a quadratic approximation, we find
\begin{equation}
\epsilon\sim m^{2}\Delta\phi^{2}\sim m^2 f^2,
\end{equation}
where $\Delta\phi\sim f$ is the approximate field separation between the two minima. The energy in the matter fluid $\rho^{\star}_{matter}$ is roughly given by the height of the deepest well, i.e.~$\rho^{\star}_{matter}\simeq\Lambda^4$. This is because at completion of the phase transition the oscillations of the scalar field span almost the whole well. Therefore, in the envelope approximation we obtain
\begin{equation}
\label{eq:eta3}
\eta \sim \kappa^{-1}.
\end{equation}
As we have mentioned, deviations from this simple picture may arise in our case. On the one hand, a certain fraction of the energy of the walls might for example be dissipated into the matter fluid. In this case, only a fraction of $\epsilon$ would lead to production of gravitational waves. This effect might be captured by the efficiency prefactor $\lambda$.\footnote{Let us also notice that bubble walls may also be generically crossed by the fluid. We neglect this effect in our discussion.}
\begin{figure}
\centering
\includegraphics[width=0.5\textwidth]{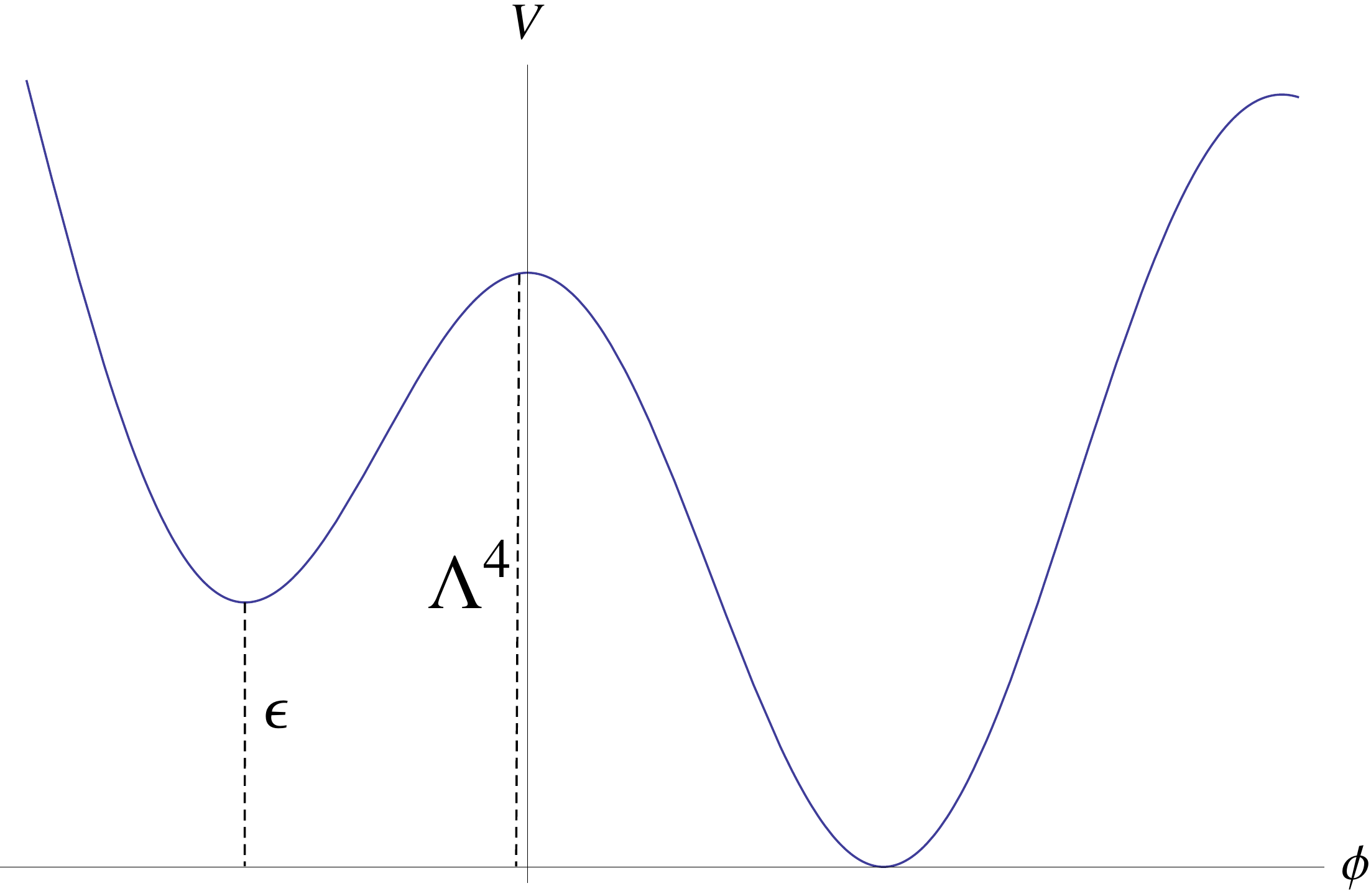}
\caption{Two-well potential. In the picture $\epsilon$ is the energy difference between the two minima, while $\Lambda^4$ is the value of the potential at the local maximum.}
\label{fig:two}
\end{figure} 
On the other hand, the energy released into the fluid while the bubbles expand and collide might also contribute to the production of gravitational waves. Namely, this energy might be converted into bulk motion of the fluid. In this case, the energy released in gravitational radiation should be larger than $\epsilon$, and could possibly be as large as $\Lambda^4$. This effect is captured by studying the fluid as a source of gravitational waves. We comment very briefly on this topic in the next subsection.

\end{subsection}

\begin{subsection}{Gravitational waves from the matter fluid}

In analogy with the case of radiation there are at least two effects which can further contribute to the total energy released in gravitational radiation during the phase transition. Here we just provide the formulae given in \cite{Caprini:2015zlo} for the thermal case, keeping in mind that they may not straightforwardly extend to our setup:
\begin{itemize}
\item[$1.$]{\textbf{Sound waves in the fluid}: this arises because a certain fraction $\lambda_{v}$ of the energy of the walls is converted after the collision into motion of the fluid (and is only later dissipated). In the case of radiation this gives a contribution
\begin{equation}
\label{eq:sw}
\frac{\rho_{GW, sw}}{\rho_{tot}}\sim \theta_{sw}\left(\frac{H_{\star}}{\delta}\right)\lambda_{v}^2\left(\frac{\eta^2}{(1+\eta)^2}\right)
\end{equation}
The prefactor $\theta_{sw}$ is expected to be smaller than $\theta$ in \eqref{eq:grojean}.}
\item[$2.$]{\textbf{Turbulence in the fluid}: one expects further contributions as a certain fraction $\lambda_{turb}$ of the energy of the walls is converted into turbulence. In the case of radiation one obtains:
\begin{equation}
\label{eq:sw}
\frac{\rho_{GW, turb}}{\rho_{tot}}\sim \theta_{turb}\left(\frac{H_{\star}}{\delta}\right)\lambda_{turb}^{3/2}\left(\frac{\eta^{3/2}}{(1+\eta)^{3/2}}\right).
\end{equation}
The prefactor is expected to be larger than $\theta$ in \eqref{eq:grojean}. Note, that for these two mechanism the dependence on $H/\delta$ is only linear.}
\end{itemize}
In certain regimes, these two effects may be larger than the one due to bubble collisions and shocks in the fluid. However, they are not fully understood, even in the case of radiation. Therefore, in the next subsection we will neglect them, and obtain only a lower bound on the relic abundance of gravitational waves. This should still be useful to understand the approximate size and frequency of the signal. Nevertheless, the reader should keep in mind that there are other possible contributions even beyond the ones mentioned in this subsection (see e.g.~\cite{Barenboim:2016mjm} for recent progress).
\end{subsection}

\begin{subsection}{Frequency and signal strength of gravitational waves}

In order to compute the relic abundance and frequency of gravitational waves emitted during the phase transition, we need to know the equation of state of the background energy density from the end of the phase transition to today. Assuming standard evolution after reheating the behaviour of the scale factor until today is essentially fixed\footnote{It is characterized by the effective number of degrees of freedom.}. Furthermore, the inflaton field generically behaves as non-relativistic matter after inflation. It remains to be addressed whether deviations from the equation of state $w\approx 0$ might occur immediately after the phase transition, before reheating. 

Due to the very large release of energy during the collision of bubble walls, it is conceivable that the fluid initially behaves relativistically. This would correspond to an early phase of radiation domination, i.e. $w=1/3$, in a similar fashion to some \emph{preheating} scenarios. Eventually, the fluid cools down and its non-relativistic behaviour is restored. Depending on the reheating temperature, this may or may not happen before the inflaton decays. If the system were in a thermal ensemble, the fluid would behave non-relativistically after $T\sim m_{\phi}\sim M$. 
For the time being, we allow for a general equation of state parameter $w$ after the phase transition and before $T\sim M$.

Therefore the background energy density at reheating is given by
\begin{equation}
\label{eq:rhorh}
\rho_{RH}=\Lambda^4\left(\frac{a_{\star}}{a_{NR}}\right)^{3(1+w)}\left(\frac{a_{NR}}{a_{RH}}\right)^{3},
\end{equation} 
where from now on the subscript $NR$ denotes that a certain quantity is evaluated at the time when the fluid becomes non-relativistic. Let us define the prefactors
\begin{align}
\nu_{w} &\equiv \left(\frac{a_{\star}}{a_{NR}}\right)\sim \left(\frac{\rho_{NR}}{\rho_{\star}}\right)^{\frac{1}{3(1+w)}},\\
\nu_{nr} &\equiv \left(\frac{a_{NR}}{a_{RH}}\right)\sim \left(\frac{\rho_{RH}}{\rho_{NR}}\right)^{1/3}.
\end{align}
The prefactor $\nu_{w}$ quantifies the duration of a period of matter domination before reheating, while $\nu_{nr}$ quantifies the duration of an early epoch of matter domination. Obviously $0<\nu_{w},\nu_{nr}\leq 1$.

The energy density in gravitational waves scales as $a^{-4}$. According to \eqref{eq:ours}, at reheating we have
\begin{equation}
\rho_{GW}(t_{RH})\approx\Lambda^4\nu_{w}^{4}\nu_{nr}^{4}\left[\theta_{0}\left(\frac{H_{\star}}{\delta}\right)^2\frac{\eta^2}{(1+\eta)^2}\right]=\nu_{w}^{-3(w-1/3)}\cdot \nu_{nr}\left[\theta_{0}\left(\frac{H_{\star}}{\delta}\right)^2\frac{\eta^2}{(1+\eta)^2}\right]\rho_{RH},
\end{equation}
where in the last step we have used \eqref{eq:rhorh}.
The relic energy density of gravitational waves is then 
\begin{equation}
\rho_{GW}(t_{0})\approx \nu_{w}^{-3(w-1/3)}\cdot\nu_{nr}\left[\theta_{0}\left(\frac{H_{\star}}{\delta}\right)^{2}\frac{\eta^2}{(1+\eta)^2}\right]\left(\frac{a_{RH}}{a_{0}}\right)^{4} \rho_{RH},
\end{equation}
where $t_{0}$ is the current age of the Universe. The ratio $a_{RH}/a_{0}$ can be determined by imposing entropy conservation. Furthermore, $\rho_{RH}$ can be computed using the standard formula for the energy density of radiation
\begin{equation}
\label{eq:rad}
\rho_{RH}=\frac{\pi^2 g_{\star}(T_{RH})}{30}T_{RH}^4.
\end{equation}
Finally, the density parameter today $\Omega_{GW}(t_{0})\equiv \frac{\rho_{GW(t_{0})}}{\rho_{crit}}=\rho_{GW(t_{0})}\frac{8\pi G_{N}}{3H_{0}^{2}}$ today is given by 
\begin{equation}
\label{eq:density}
\Omega_{GW}(t_{0})\simeq \frac{10^{-5}\nu_{w}^{-3(w-1/3)}\nu_{nr}}{h^{2}}\cdot \theta_{0}\left[\frac{10^{2}}{g_{*}(T_{RH})}\right]^{1/3}\cdot\left[\frac{H_{\star}}{\delta}\right]^{2}\cdot\frac{\eta^2}{(1+\eta)^2},
\end{equation}
where $h\equiv H_{0}/(100 \text{~km}\cdot\text{s}^{-1}\cdot\text{Mpc}^{-1})$.

Let us now estimate the peak frequency of the emitted radiation. For this quantity the only relevant parameter is $\delta$. Frequencies scale as $\sim a^{-1}$, therefore we have: 
\begin{equation}
\label{eq:frequency}
\omega_{0}\sim\omega_{peak}\left(\frac{a_{\star}}{a_{NR}}\right)\left(\frac{a_{NR}}{a_{RH}}\right)\left(\frac{a_{RH}}{a_{0}}\right)\sim \omega_{peak}\cdot \nu_{w}\cdot \nu_{nr}\cdot \left(\frac{a_{RH}}{a_{0}}\right).
\end{equation}
By combining \eqref{eq:frequency}, \eqref{eq:frequencypeak}, $H^{2}\sim \rho/M_{p}^2$ and \eqref{eq:rad} one obtains 
\begin{equation}
\omega_{0}\sim  10^{8} \text{Hz}\cdot  \sigma\cdot\nu_{w}\cdot\nu_{nr}\left(\frac{\delta}{H_{*}}\right)\left(\frac{g_{*}(T_{RH})}{10^{2}}\right)^{1/6}\left[\frac{T_{RH}}{10^{15}\text{GeV}}\right].
\end{equation}
We have therefore determined the relevant parameters of the emitted radiation as function of $\delta$ and $\eta$. Now we can plug in \eqref{eq:delta} and \eqref{eq:eta3}, to obtain final formulae. Then we have:
\begin{align}
\Omega_{GW}(t_{0})h^2&\simeq 10^{-5}\nu_{w}^{-3(w-\frac{1}{3})}\cdot\nu_{nr}\cdot \theta_{0}\left[\frac{10^{2}}{g_{*}(T_{RH})}\right]^{1/3}\kappa^{-2}\\
\omega_{0}&\simeq 10^{8} \text{Hz}\cdot \sigma\cdot\nu_{w}\cdot\nu_{nr}\cdot \left(\frac{g_{*}(T_{RH})}{10^{2}}\right)^{1/6}\left[\frac{T_{RH}}{10^{15}\text{GeV}}\right].
\end{align}
In the envelope approximation, it is also possible to compute the full spectrum of the gravitational radiation emitted from the collision of the bubble walls. This reads \cite{Caprini:2015zlo}:
\begin{align}
\label{eq:omegagw}
\Omega_{GW}(t_{0})h^2&\simeq 10^{-5} \nu_{w}^{-3(w-1/3)}\nu_{nr}\cdot \theta_{0}\cdot \kappa^{-2}\left[\frac{10^{2}}{g_{*}(T_{RH})}\right]^{1/3} S_{env}(\omega)\\
\nonumber \text{with}\quad S_{env}(\omega)&=\frac{3.8(\omega/\omega_{0})^{2.8}}{1+2.8(\omega/\omega_{0})^{3.8}}.
\end{align}
In order to estimate the maximal possible size of our signal, let us now specify to the case in which the energy present after the bubble collisions is converted into radiation. This corresponds to setting $w=1/3$ in \eqref{eq:omegagw}. At $T\sim M$ the inflaton goes back to a non-relativistic behaviour. The prefactors $\nu_{w}$ and $\nu_{nr}$ can now be explicitly computed, using $\rho_{NR}\sim T_{NR}^{4}\sim M^{4}, \rho_{RH}\sim T_{RH}^4$. We obtain
\begin{align}
\nu_{w} &=\left(\frac{\rho_{NR}}{\rho_{\star}}\right)^{1/4}\sim \frac{M}{\Lambda}\sim \frac{\kappa^{1/4}m^{1/2}}{f^{1/2}}\\
\label{eq:nr}
\nu_{nr} &=\left(\frac{\rho_{RH}}{\rho_{NR}}\right)^{1/3}\sim \left(\frac{T_{RH}}{m\kappa^{1/2}}\right)^{4/3}.
\end{align}
From \eqref{eq:nr}, it is clear that the largest signal is obtained for $T_{RH}\sim T_{NR}\sim M$. In this case $\nu_{nr}\sim 1$ and the signal is completely unsuppressed. For completeness, let us mention that the largest suppression of the signal occurs for $w\approx 0$. In this case, as should be clear from \eqref{eq:rhorh}, the background energy density scales as matter from completion of the phase transition until reheating.

In Fig.~\ref{fig:sens} we plot the spectrum \eqref{eq:omegagw} for three different choices of parameters, using $w=1/3$. We fix $m\simeq 10^{-5}M_{p}$ as required by observations. We have chosen parameters in such a way as to maximize the overlap with sensitivity regions of current and future space- and ground-based detectors, which are bounded by dashed lines in the plot. We have also fixed $\theta_{0}=10^{-2}, \sigma=10^{-1}$. Our plot provides examples of the wide range of frequencies that can be obtained in our setting, simply varying the axion decay constant, the reheating temperature and the number of local minima. Interestingly, for reasonable choices of parameters the signals are in the reach of future detectors.

\begin{figure}
\centering
\includegraphics[width=0.7\textwidth]{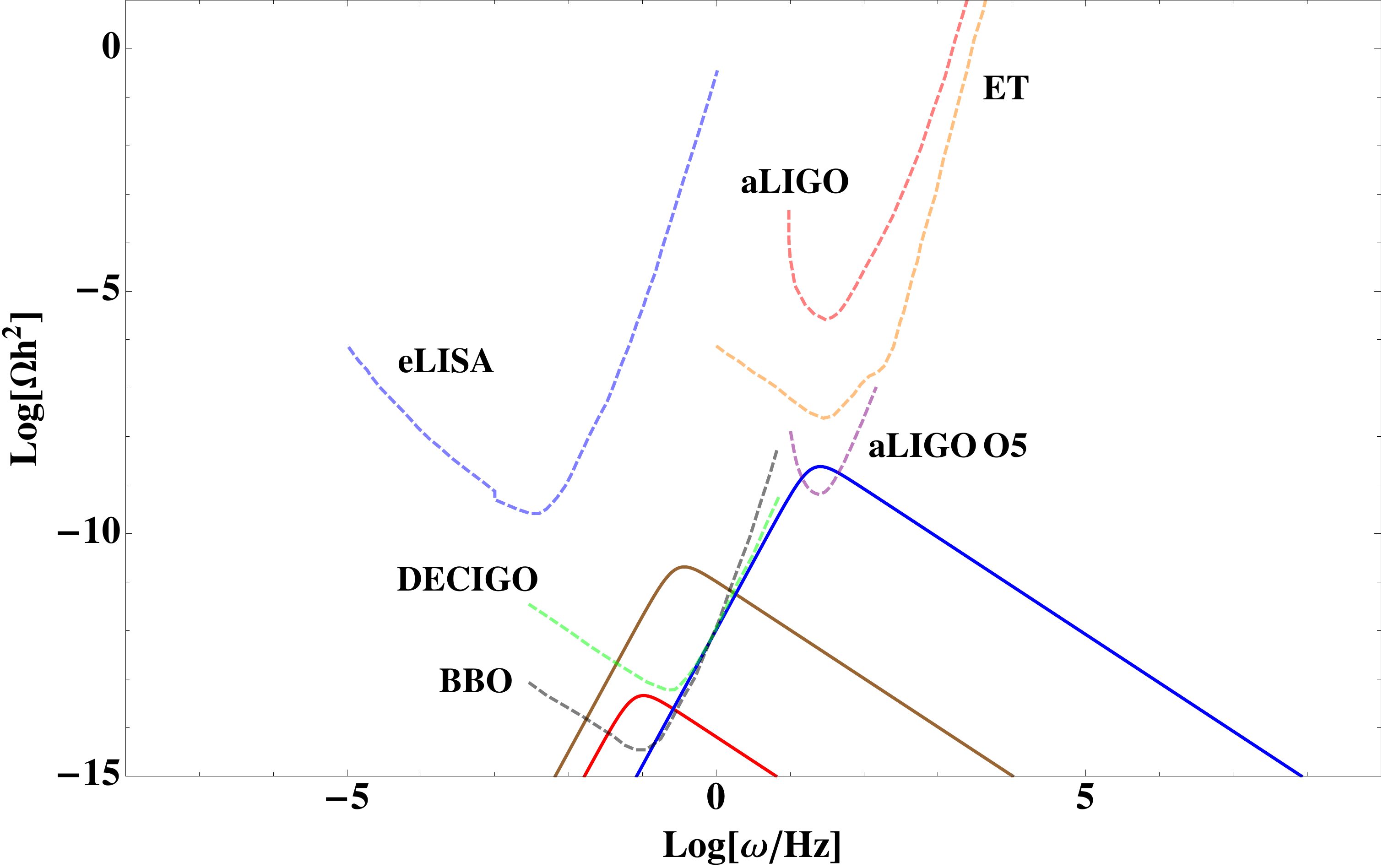}
\caption{Gravitational wave spectra as in \eqref{eq:omegagw} with $w=1/3$. The inflaton mass is fixed to $m\sim 10^{-5}M_{p}$. Spectra are shown as solid lines for different values of $\kappa, f$ and $T_{RH}$: the blue curve is obtained for $\kappa=5, f=0.1 M_{p}, T_{RH}\sim 10^{12}$ GeV; the brown curve for $\kappa=10, f=0.01 M_{p}, T_{RH}\sim 10^{11}$ GeV; the red one for $\kappa=70, f=0.001 M_{p}, T_{RH}\sim 10^{11}$ GeV. We have also taken $w=1/3, \theta_{0}=10^{-2}, \sigma=10^{-1}$ in \eqref{eq:omegagw}. For the values of the reheating temperature considered here, we have $g_{*}(T_{RH})\sim 10^2$. Sensitivity curves of some ground- and space-based interferometers are shown for comparison as dashed curves (data taken from \cite{Moore:2014lga}).}
\label{fig:sens}
\end{figure}

\end{subsection}
\end{section}

\begin{section}{Conclusions}
\label{sec:conclusions}

In this paper we investigated the production of gravitational waves from post-inflationary dynamics in models of Axion Monodromy inflation. We expect such phenomena to also occur for generic axionic fields with potentials with sizable modulations, albeit the numerical results may differ significantly in these cases.

The main observation is that in models of axion monodromy inflation, the inflaton potential consists of a monotonic polynomial with superimposed cosine-modulations. While these modulations have to be small to allow for successful inflation, they tend to dominate near the bottom of the potential. In fact, these cosine `wiggles' can be large enough such that the potential exhibits a series of local minima. 

After inflation ends, the inflaton is exploring this wiggly part of the potential. As a consequence of Hubble friction, the rolling axion can get stuck in one of the wells. This may happen well before the inflaton reheats the standard model degrees of freedom. Since the energy density in the axion field decreases together with $H$, the field is more likely to get caught in one of the last minima. We therefore focused on a two-well setting, which is obtained by ``zooming'' into the full monodromy potential.

Taking fluctuations of the inflaton field into account, the inflaton field does not necessarily get caught in one unique local minimum in the entire Hubble patch. If field fluctuations are sufficiently large, a phase decomposition occurs such that at least two different vacua are populated after inflation. The probability of this occurring is given by $\mathcal{P}\sim \delta\rho/\Delta\rho$,  where $\delta\rho$ is the uncertainty in the axion energy density induced by the field fluctuations and $\Delta\rho$ is the frictional loss of energy in one oscillation.

We can distinguish two sources of field fluctuations which may lead to a phase decomposition. Firstly, there are the classical inhomogeneities naturally inherited from inflation. Secondly, there are the intrinsic quantum uncertainties characterising any quantum field. These two sources are essentially indistinguishable at very early and late times. However they are in principle of different size in the intermediate regime that we are interested in. In particular, we have found that inflationary fluctuations are more likely to induce a phase decomposition: the probability that they do so is $\mathcal{P}\sim \kappa^{-1/3}(m/M_{p})(M_{p}/f)^{5/3}$. Here $m$ is the mass of the inflaton-axion, $f$ the axion decay constant which defines the axion periodicity $2 \pi f$ and $\kappa/\pi$ roughly counts the number of local minima of the potential. Therefore, we observe that a phase decomposition is likely for $\kappa\sim O(10)$ and $f\lesssim 0.3\cdot 10^{-2} M_{p}$. The probability that quantum fluctuations induce a phase decomposition is smaller by a factor $(f/(M_{p}\kappa))^{-4/3}m/M_{p}$.

Furthermore, due to the oscillatory term in the axion potential, fluctuations can experience exponential growth. This effect arises generically for modes with $k\sim m$ at the time when the field is rolling over the last wells. These fluctuations never exited the horizon, therefore they are effectively quantum modes. We extended the study of enhancement in \cite{Jaeckel:2016qjp} to the case with varying $H$, but still neglecting the gravitational field in the equations of motion. Numerically and using a classical approximation, we observe the existence of a region of parameter space where a phase decomposition is likely to occur. This happens roughly in the same regime where inflationary classical fluctuations are also likely to induce a phase decomposition. However, given the exponential enhancement of quantum modes, the latter are more likely to be the dominant cause of phase decomposition. Enhanced quantum modes quickly become classical, therefore we expect our main results to hold even after a more detailed analysis, which we leave for future work.
For larger values of $f$ modes are not enhanced. Phase decomposition, although unlikely, might still occur as a result of quantum or classical fluctuations. For smaller decay constants fluctuations are very strongly enhanced. Phase decomposition occurs but it is hard to understand the physics in such a highly non-linear regime. 

If a phase decomposition occurs, bubbles containing minima of lowest energy expand. Collisions of these bubbles source gravitational waves. We estimated the energy density and frequency of the emitted radiation in terms of the axion parameters, in the envelope approximation. Furthermore, we note that the matter fluid associated to the oscillating inflaton may also radiate gravitational waves. This is similar to the case of a thermal phase transition. The spectrum of the emitted radiation can peak in a wide range of frequencies (from mHz to GHz), depending on the reheating temperature and on the time of the phase transition. In this sense, our source is similar to other post-inflationary phenomena, such as preheating and cosmic strings. However, it is interesting to observe that the frequency may be lowered in our case since a matter dominated phase can follow the phase transition. The spectrum is at least partially in the ballpark of future space- and ground-based detectors.
Thus, we can hope that axion monodromy may one day be investigated by means of gravitational wave astronomy.

\end{section}

\section*{Acknowledgments}

We thank Robert Brandenberger, Christophe Grojean, David Hindmarsh, Sascha Leonhardt, Patrick Mangat, Viraf Mehta, Michael Schmidt, Michael Spannowsky and Alexander Westphal for useful discussions. This work was partly supported by the DFG Transregional Collaborative Research Centre TRR 33 “The Dark Universe". F.R. is supported by the DFG Graduiertenkolleg GRK 1940 “Physics Beyond the Standard Model”.

\begin{appendices}

\begin{section}{Scalar field fluctuations after inflation}
\label{sec:after}

In this appendix, we provide a detailed analysis of the evolution of scalar field fluctuations after inflation. We begin by deriving the Klein-Gordon equations for a scalar field and its fluctuations with potential \eqref{eq:potential}, including the gravitational field. We then focus on the simple case of a purely quadratic potential. This gives us the opportunity to review why scalar field fluctuations behave like dark matter perturbations. We then provide equations to study the fluctuations in the full potential containing the `wiggles'.

\begin{subsection}{Equations of motion}

Let us begin with the equations of motion of a scalar field in the post-inflationary universe. The starting point is the Klein-Gordon equation:
\begin{equation}
\label{eq:kg}
\frac{1}{\sqrt{-g}}\partial_{\mu}\left[g^{\mu\nu}\sqrt{-g}\partial_{\nu}\phi \right]+V^{'}(\phi)=0.
\end{equation}
The metric appearing in \eqref{eq:kg} is the perturbed FRW metric (here we follow \cite{Baumann}):
\begin{equation}
\label{eq:metric}
ds^{2}=a^{2}(\tau)\left[(1+2A)d\tau^{2}-2B_{i}dx^{i}d\tau-(\delta_{ij}+h_{ij})dx^{i}dx^{j}\right],
\end{equation}
where $B_{i}=\partial_{i}B+\hat{B}_{i}$ and the hat denotes divergenceless vectors and traceless tensors. In particular, we can consistently focus only on scalar modes, since the vectors can be gauged away and the tensors are not sourced by $\delta\phi$. We therefore have
\begin{equation}
h_{ij}^{scalar}=2C\delta_{ij}+2\partial_{<i}\partial_{j>}E,
\end{equation}
where $\partial_{<i}\partial_{j>}E\equiv \left[\partial_{i}\partial_{j}-(1/3) \delta_{ij}\nabla^2\right]E$.
Perturbations defined by $\phi(t,\mathbf{x})=\phi_{0}(t)+\delta\phi(t,\mathbf{x})$ and \eqref{eq:metric} are not gauge invariant. However, the following quantities are gauge invariant:
\begin{align}
\label{eq:gaugeinv}
\nonumber \Psi &\equiv A+\mathcal{H}(B-E^{'})+(B-E^{'})^{'}\\
\nonumber \Phi &\equiv -C-\mathcal{H}(B-E^{'})+\frac{1}{3}\nabla^{2}E\\
\nonumber \overline{\delta\phi} &\equiv \delta\phi-\phi_{0}^{'}(B-E^{'}).
\end{align}
In what follows we will perform our computation in the \emph{Newtonian gauge}, defined by:
\begin{equation}
\label{eq:newton}
B=E=0, C=-\Phi.
\end{equation}
In the latter, the perturbed metric reads:
\begin{equation}
\label{eq:newtoniangauge}
ds^{2}=[(1+2\Psi)dt^{2}-(1-2\Phi)a^{2}\delta_{ij}dx^{i}dx^{j}].
\end{equation}
The components $\Phi$ and $\Psi$ are related by the perturbed Einstein equations. In the absence of off-diagonal terms in the spatial components of the perturbed stress-energy tensor the Einstein equations impose $\Phi=\Psi$. With this constraint, the metric \eqref{eq:newtoniangauge} provides the newtonian limit of general relativity. 

We are now in a position to write down the Klein-Gordon equation \eqref{eq:kg}, expanding the scalar field as $\phi=\phi_{0}+\delta\phi$ and keeping only the leading order terms in the perturbed quantities $\delta\phi, \Phi$. Then the background $\phi_{0}$ obeys:
\begin{equation}
\label{eq:backgroundkg}
\ddot{\phi}_{0}+3\frac{\dot{a}}{a}\dot{\phi}_{0}+V^{'}(\phi_{0})=0,
\end{equation}
while the inhomogeneous $\delta\phi(t,\mathbf{x})$ satisfies:
\begin{equation}
\label{eq:fluctuationskg}
\ddot{\delta\phi}+3\frac{\dot{a}}{a}\dot{\delta\phi}+\left(V^{''}(\phi_{0})-\frac{\nabla^2}{a^2}\right)\delta\phi-4\dot{\phi}_{0}\dot{\Phi}+2V^{'}(\phi_{0})\Phi=0.
\end{equation}
Furthermore, there are three Einstein equations relating the gravitational field $\Phi$ to the fluctuation $\delta\phi$:
\begin{align}
\label{eq:constrainte}
\dot{\Phi}+\frac{\dot{a}}{a}\Phi&=4\pi G\dot{\phi}_{0}\delta\phi\\
\label{eq:poissone}
\frac{\Delta\Phi}{a^{2}}-3\frac{\dot{a}}{a}\dot{\Phi}-3\frac{\ddot{a}^{2}}{a^{2}}&=4\pi G\delta\rho\\
\ddot{\Phi}+4\frac{\dot{a}}{a}\dot{\Phi}+(2H^{2}+H^2)\Phi&=4\pi G \left(\dot{\phi}_{0}\dot{\delta\phi}+V^{'}(\phi_{0})\right)\delta\phi,
\end{align}
where $\delta\rho=\dot{\phi}_{0}\dot{\delta\phi}+V^{'}(\phi_{0})$ is the perturbed energy density. The Einstein equations \eqref{eq:constrainte} and \eqref{eq:poissone} lead to a Poisson equation with a gauge invariant energy density on the right hand side. In total, we have five equations for three quantities and therefore it is enough to consider only one of the Einstein equations.

\end{subsection}

\begin{subsection}{Scalar field as Dark Matter}

In order to understand why perturbations of a scalar field behave like dark matter fluctuations, we now focus on the case of a purely quadratic potential $V(\phi)=\frac{1}{2}m^{2}\phi^{2}$. In this case it can be shown that fluctuations obeying \eqref{eq:fluctuationskg} behave like an ideal pressureless fluid \cite{Ratra:1990me, Gorbunov:2011zzc}. The strategy is as follows \cite{Gorbunov:2011zzc}: by using a  WKB ansatz for $\phi_{0}$ and $\delta\phi$ one can define three fluid quantities for a scalar field: energy density, velocity potential and pressure. One then performs an expansion of \eqref{eq:fluctuationskg} in powers of $(m^{-1})$. The leading order in this expansion corresponds to the Euler equation for a perfect pressureless fluid. The subleading order gives the continuity equation. This establishes a dictionary between scalar field and fluid quantities. More precisely, the correspondence involves field quantities which are averaged over one period $T\sim m^{-1}$. One can then use the standard fluid approach to prove that scalar field energy fluctuations grow like $a$ on subhorizon scales. 

We now briefly review the approach of \cite{Gorbunov:2011zzc}: the aim is to provide equations which can be extended to include non-linearities. The following analysis is valid in the regime $m\gg H, k/a \gg H, k^2/(a^2 H)\ll m$. The first two conditions are satisfied in our setup. However, this is not necessarily the case for the third one. Indeed, as explained in Sec. \ref{sec:phases}, we are mainly interested in modes with $k\gtrsim m$ at horizon re-entry, as these are most efficiently leading to a phase decomposition. However, due to the expansion of the universe $k^2/(a^2 H)$ decreases as $a^{-1/2}$. Therefore, even if the condition  $k^2/(a^2 H)\ll m$ is originally not satisfied, it rapidly becomes fulfilled.

The starting point of the analysis is a WKB ansatz for the background and for the fluctuations (see also \cite{Urena-Lopez:2013naa, Braaten:2016kzc} for related discussions):
\begin{align}
\label{eq:wkbbackground}
\phi_{0}&=u(t)\cos(mt)+w(t)\sin(mt)\\
\label{eq:wkbflucts}
\delta\phi &= B\left(t,\mathbf{x}\right)\sin(mt)+A\left(t,\mathbf{x}\right)\cos(mt).
\end{align}
The crucial point is that $u,w,B$ and $A$ vary on time scales which are much longer than $m^{-1}$. Using \eqref{eq:wkbbackground} in \eqref{eq:backgroundkg} and the Friedmann equation, we find $u\sim (mt)^{-1}$. One can then check that $w$ represents a subleading correction, of $O(m^{-2})$. Assuming $\Phi\sim O(m^{0})$, the choice consistent with Einstein equations is $B\sim O(m^{0})$ and $A\sim O(m^{-1})$. At leading order in $m^{-1}$, i.e. at $O(m^{1})$, \eqref{eq:wkbflucts} reads:
\begin{equation}
\label{eq:m1}
\dot{B}+\frac{3}{2}\frac{\dot{a}}{a}B+mu\Phi=0.
\end{equation}
Let us now write down the relevant fluid quantities. In analogy with the stress-energy tensor of a perfect fluid, they are defined as:
\begin{align}
\label{eq:fluid}
\delta\rho &= \overline{\delta T_{00}}\\
v_{i} &=-\frac{1}{\overline{(\rho+p)}} \overline{\delta T^{0}_{i}}\\
\delta p &= - \overline{\delta T^{i}_{j}},
\end{align}
where the overlines denote an average over a period $m^{-1}$. In what follows we will use the velocity potential $v$, defined as $v_{i}= \partial_{i} v$. Computing explicitly the components of the perturbed stress energy tensor, using \eqref{eq:wkbbackground},\eqref{eq:wkbflucts},\eqref{eq:m1} one finds the leading order expressions for the fluid quantities \eqref{eq:fluid}:
\begin{align}
\label{eq:fluid2}
\delta\rho &= m^{2}\left[uA+wB \right]\\
\label{eq:vel}
v & = \frac{B}{mu a}\\
\label{eq:pressureless}
\delta p &= 0.
\end{align}
As expected, we see that the effective pressure of the fluctuations vanishes at leading order.

A perfect pressureless newtonian fluid in an expanding background is subject to the familiar continuity and Euler equations (see e.g. \cite{Baumann}):
\begin{align}
\label{eq:continuity}
\dot{\delta}&=-\frac{1}{a}\nabla\cdot \mathbf{v}\\
\label{eq:euler}
\dot{\mathbf{v}}+H\mathbf{v} &=-\frac{1}{a}\nabla\Phi,
\end{align}
where $\delta\equiv \delta\rho/\rho$ and $v_{i}\equiv \partial_{i} v$. According to \eqref{eq:vel}, $v\sim B$. We see therefore that \eqref{eq:m1} closely resembles the Euler equation \eqref{eq:euler}. In fact, it can be shown that \eqref{eq:m1}, rewritten in terms of $v$, is equivalent to the relativistic generalisation of \eqref{eq:euler}. 

The next-to-leading order of \eqref{eq:wkbflucts}, i.e. the $O(m^{0})$, is obtained by taking into account that both the scale factor and the gravitational potential have oscillatory subleading terms, $\Phi_{osc}\sim O(m^{-1})$ and $a_{osc}\sim O(m^{-2})$. The details can be found in \cite{Gorbunov:2011zzc}. Here we report only the final result:
\begin{equation}
\label{eq:m0}
\ddot{B}+3\frac{\dot{a}}{a}\dot{B}-\frac{\nabla^2}{a^2}B-2m\dot{A}-3m\frac{\dot{a}}{a}A+4mu\dot{\Phi}+2m^{2}w\Phi-\frac{3\pi}{2} Gm^{2}u^{2}B=0.
\end{equation}
Since $A, B$ and $\Phi$ vary on time scales which are larger than $m^{-1}$, the first two terms in \eqref{eq:m0} are subleading compared to the other terms in the equation. The last two terms come from the subleading parts of the background and the scale factor. Neglecting them, we see that \eqref{eq:m0} closely resembles \eqref{eq:continuity}. More precisely, it can be shown that the full \eqref{eq:m0}, rewritten in terms of $v$ and $\delta\rho$, reproduces the relativistic generalisation of \eqref{eq:continuity}.

The discussion of this section therefore proves that scalar field fluctuations behave like an ideal pressureless fluid after inflation. Having established this equivalence, one can show that $\delta\rho/\rho\sim a$ on subhorizon scales, using \eqref{eq:continuity}, \eqref{eq:euler} and the Poisson equation for the gravitational field (see e.g. \cite{Baumann}).

\end{subsection}

\begin{subsection}{Equation including non-linearities}
\label{sub:nonlinearities}

The next goal is to write down the analogue of \eqref{eq:m1} and \eqref{eq:m0} for the axion monodromy potential \eqref{eq:potential}. This amounts to including non-linearities in the fluid equations. It turns out that in this case the scalar field perturbations do not behave like a pressureless fluid. Rather, the pressure is small but non-vanishing, and correspondingly the fluid is characterised by a small sound speed. In our setup we have $\kappa\gtrsim O(1)$, therefore $\Lambda^4/f^2\sim O(m^{2})$ and the $O(m^{2})$ expansion of \eqref{eq:fluctuationskg} is not necessarily trivially satisfied. One would therefore have to assign a certain order to the new terms involving the fast oscillations with argument $\phi_{0}/f$, such that a consistent solution to Einstein equations and the equations of motion can be found. 

In this subsection, we take a different approach. Namely, we derive two equations from \eqref{eq:fluctuationskg} by splitting the sine and cosine oscillations. We then obtain equations that involve terms of different orders in $m^{-1}$. The three relevant orders are $O(m^{2}), O(m^{1}), O(m^{0})$. At these orders there is no need to consider higher harmonics in $mt$, therefore we only keep terms in \eqref{eq:fluctuationskg} that are either non-oscillating or oscillating with frequency $m^{-1}$.
The result of such a computation is the following system of equations for $A$ and $B$:
\begin{align}
\label{eq:sin}
&\ddot{B}-2m\dot{A}+3\frac{\dot{a}}{a}B-B\frac{\Lambda^4}{f^2}\cos\left(\frac{\phi_{0}}{f}\right)-\frac{\nabla^{2}}{a^{2}}B +4\dot{\Phi} m u - 4\left[-\frac{\dot{w}}{2}+m^{2}\pi G u^{2} B \right] \\
\nonumber &+ 2\left[m^{2}w-\frac{\Lambda^4}{f}\sin\left(\frac{\phi_{0}}{f}\right)\right]\Phi-\frac{3}{4}m\pi G A\dot{u}^{2} -2m^3\pi G  wu B=0.
\end{align}

\begin{align}
\label{eq:cos}
&\ddot{A}+2m\dot{B}+3\frac{\dot{a}}{a}A-A\frac{\Lambda^4}{f^2}\cos\left(\frac{\phi_{0}}{f}\right)-\frac{\nabla^{2}}{a^{2}}A-4\dot{\Phi} m w-4\left[\frac{\dot{u}}{2}+m^{2}\pi G uw B\right]\\
\nonumber &+ 2\left[m^{2}u-\frac{\Lambda^4}{f}\sin\left(\frac{\phi_{0}}{f}\right)\right]\Phi-\frac{3}{4}\pi G m B\dot{u}^{2}+2m^{3}\pi Gu^{2} B=0.
\end{align}
The next step to analyse these equations is to expand $A$ and $B$ in Fourier modes in $k$. Notice that \eqref{eq:sin} and \eqref{eq:cos} contain oscillatory terms with argument $\phi_{0}/f$. Those terms are the cause of potential resonant behaviour of the modes $B_{k}$ and $A_{k}$. Solutions to \eqref{eq:sin}, \eqref{eq:cos} should be sought numerically. While we leave a detailed study for future work, let us notice that before performing this numerical study one has to properly average and Fourier expand the functions $\cos(\phi_{0}/f), \sin(\phi_{0}/f)$.

\end{subsection}

\end{section}

\begin{section}{Tuned small field inflation}
\label{sec:inflection}

In this appendix we would like to briefly address an issue related to our setup. In principle, while oscillating along the full potential \eqref{eq:potential}, the inflaton may come close to a local maximum, without being able to actually reach it as a consequence of friction. In this case, the field motion may satisfy the slow roll conditions, therefore leading to a further inflationary phase. If the local maximum is flat enough, we might be in a scenario which closely resemble that of inflation at an inflection point (see \cite{Baumann:2014nda} and references therein, see also \cite{Kobayashi:2014ooa}, \cite{Choi:2016eif},\cite{Parameswaran:2016qqq, Kadota:2016jlw} for related previous work).
If inflation is viable in this setup, then current observational constraints on $m$ may be relaxed. Indeed in this case the axion mass would not necessarily be proportional to the inflationary power spectrum. This may turn out to be particularly relevant for the signatures that we described in this paper, as the probability of having a phase decomposition is suppressed by the smallness of $m$.
Let us also remark that the potentials that we will consider in this Appendix may be relevant also for non-monodromic scenarios with more than one axion, such as alignment and winding models \cite{Kim:2004rp, Berg:2009tg, Ben-Dayan:2014zsa}.

Let us then start by assuming that the inflaton slowly approaches a stationary inflection point $\phi_{*}$ of the potential. We first estimate how many e-foldings would be generated as the field comes close to $\phi_{*}$. In this Appendix we work in Planck units, i.e. we set $M_{p}\equiv 1$.

Around an inflection point, the potential can be approximated as
\begin{equation}
\label{eq:infpot}
V=V_{0}\left(1+\gamma\frac{\phi^{3}}{3}\right).
\end{equation}
The slow roll parameters are therefore given by
\begin{equation}
\label{eq:slowroll}
\epsilon=\frac{1}{2}\gamma^{2}\phi^{4},\quad \eta=2\gamma\phi.
\end{equation}
The number of e-foldings can be estimated as:
\begin{align}
\label{eq:efolds}
\mathcal{N}&=\int\frac{d\phi}{\sqrt{2\epsilon}}=\frac{1}{\gamma}\int_{\phi_{e}}^{\phi_{*}}\frac{d\phi}{\phi^{2}}\\
\simeq \frac{1}{\gamma\phi_{*}}\simeq \frac{2}{|\eta|}.
\end{align}
The latter equation may represent a conflict with observations: indeed in this model $\epsilon$ is very small, i.e. $\epsilon\ll 10^{-2}$, therefore we need $\eta\sim 10^{-2}$ to satisfy the observational constraint on $n_{s}-1=2\epsilon-6\eta\simeq 3\cdot 10^{-2}$. From \eqref{eq:efolds}, we obtain $\mathcal{N}\sim 10^{2}$. Therefore it seems that in this simple case we obtain too many e-foldings.

However it is very unlikely that the field gets caught around such an inflection point. Indeed, the latter is the point after which there are no more local minima of the potential \eqref{eq:potential}, and therefore sits far from the minimum of the parabola. As we have already remarked, the field is more likely to get caught in one of the minima at the bottom of the potential. Therefore, we now consider a more generic stationary point and modify the potential \eqref{eq:infpot} by including also a small quadratic contribution. 

We focus on the following expansion around a stationary point $\phi_{0}$:
\begin{equation}
\label{eq:modpot}
V=V_{0}\left[1+\frac{\alpha}{2}(\phi-\phi_{0})^{2}+\frac{\beta}{3!}(\phi-\phi_{0})^{3}\right],
\end{equation}
Here we take $\alpha$ and $\beta$ to be positive without loss of generality. We want to assess the feasibility of such an inflationary potential (see \cite{Choi:2016eif} for the case in which a linear term is included, instead of a quadratic term). We take $\alpha\ll \beta$, and focus on small field ranges, so that $|\phi-\phi_{0}|<1$ all along the inflationary trajectory.

The slow roll parameters are:
\begin{align}
\label{eq:slowr}
\nonumber \epsilon &\equiv\frac{1}{2}\left[\frac{V'}{V}\right]^{2}\simeq\frac{1}{2}\left[\alpha(\phi-\phi_{0})+\frac{\beta}{2}(\phi-\phi_{0})^{2}\right]^{2}\\
\eta &\equiv\frac{V^{''}}{V}\simeq \alpha+\beta(\phi-\phi_{0})\Rightarrow \phi-\phi_{0}=\frac{\eta-\alpha}{\beta}.
\end{align}
The number of efoldings is given by:
\begin{equation}
\label{eq:numbe}
\mathcal{N}_{\star}=\int_{\phi_{e}}^{\phi_{\star}}\frac{d\phi}{\sqrt{2\epsilon}}=\frac{1}{\alpha}\ln\left[\frac{(\eta_{\star}-\alpha)(\alpha+\frac{\eta_{e}-\alpha}{2})}{(\eta_e-\alpha)(\alpha+\frac{\eta_\star-\alpha}{2})}\right],
\end{equation}
where $\phi_{\star}$ is the field value at the beginning of inflation.
It is convenient to express $\eta_{\star}$ in terms of $\mathcal{N}_{\star}$, by inverting \eqref{eq:numbe}:
\begin{equation}
\label{eq:etastar}
\eta_{\star}=\frac{e^{\alpha\mathcal{N}_{\star}}\alpha(\eta_{e}-\alpha)}{2\left[\alpha+\frac{\eta_{e}-\alpha}{2}(1-e^{\alpha\mathcal{N}_{\star}})\right]}\simeq -\frac{\eta_{e}}{-2+\mathcal{N}_{\star}\eta_{e}}-\frac{\alpha}{2},
\end{equation}
where in the last step we have kept only the first order in $\alpha$.
Similarly, we can express $\epsilon_\star$ in terms of $\eta_{\star}$, then in terms of $\mathcal{N}_{\star}$ by means of \eqref{eq:etastar}:
\begin{equation}
\epsilon_\star\simeq \frac{\eta_{e}^3}{4\beta^2(-2+\eta_e\mathcal{N}_{\star})^3}\left[\alpha+\frac{\eta_e}{2(-2+\eta_e\mathcal{N}_{\star})}\right].
\end{equation}
The end of inflation is determined either by the condition $\epsilon_e=1$ or by $\eta_e=-1$. In our case, we find:
\begin{align}
\nonumber \epsilon_e &=1\Rightarrow |\phi^\epsilon_e-\phi_0|=\frac{2^{3/4}}{\beta^{1/2}}+\frac{\alpha}{\beta}\\
\eta_e &=-1\Rightarrow |\phi_{e}^{\eta}-\phi_0|=\frac{\alpha}{\beta},
\end{align}
therefore, under the assumption $\alpha\ll \beta$, $\eta_e=-1$ determines the end of inflation. Notice also that the ratio $\alpha/\beta$ sets precisely the inflationary range. 
Now we can write down the spectral index and the tensor-to-scalar ratio in terms of $\alpha,\beta,\mathcal{N}_{\star}$:
\begin{align}
n_{s}&=1-6\epsilon_\star+2\eta_\star\simeq 1-\frac{1}{2+\mathcal{N}_\star}\left[2+\frac{3}{4\beta^2(2+\mathcal{N}_\star)^3}\right]-\alpha\left[1+\frac{2}{3\beta^2(2+\mathcal{N}_\star)^3}\right]\\
r&=16\epsilon_\star\simeq \frac{4}{\beta^2(2+\mathcal{N}_\star)^3}\left[\alpha+\frac{1}{2(2+\mathcal{N}_\star)}\right].
\end{align}
We are therefore ready to study the parameter space of the model described by \eqref{eq:modpot}. First, we fix $\mathcal{N}_\star=50$ and plot the constraints on the parameters $\alpha$ and $\beta$ obtained by imposing the observed values for $n_s$ and $r$: $n_s=0.9652\pm 0.0047$, $r<0.10$ at $95 \% $ C.L. \cite{Ade:2015xua}. We then repeat the analysis for $\mathcal{N}_{\star}=60$.

We present the results in Fig. (\ref{fig:N50}) and (\ref{fig:N60}). We see that a model based on \eqref{eq:modpot} is still a viable candidate to explain the CMB observables.

\begin{figure}[h]
    \centering
    \begin{minipage}{.5\textwidth}
        \centering
        \includegraphics[scale=0.5]{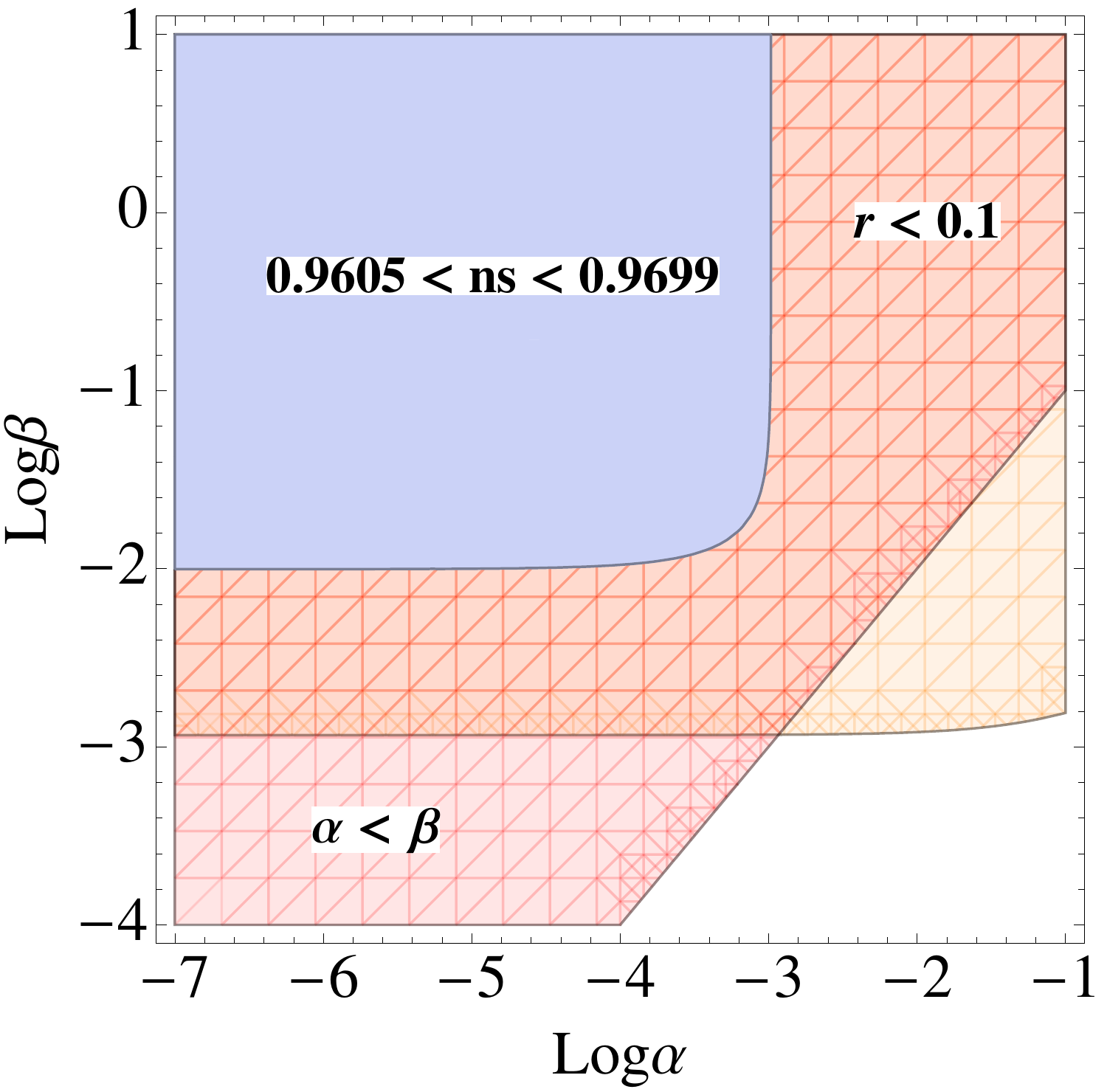}
        \caption{Constraints on the parameters $\alpha$ and $\beta$ for $\mathcal{N}_{\star}=50$.}
        \label{fig:N50}
    \end{minipage}%
    \begin{minipage}{0.5\textwidth}
        \centering
        \includegraphics[scale=0.5]{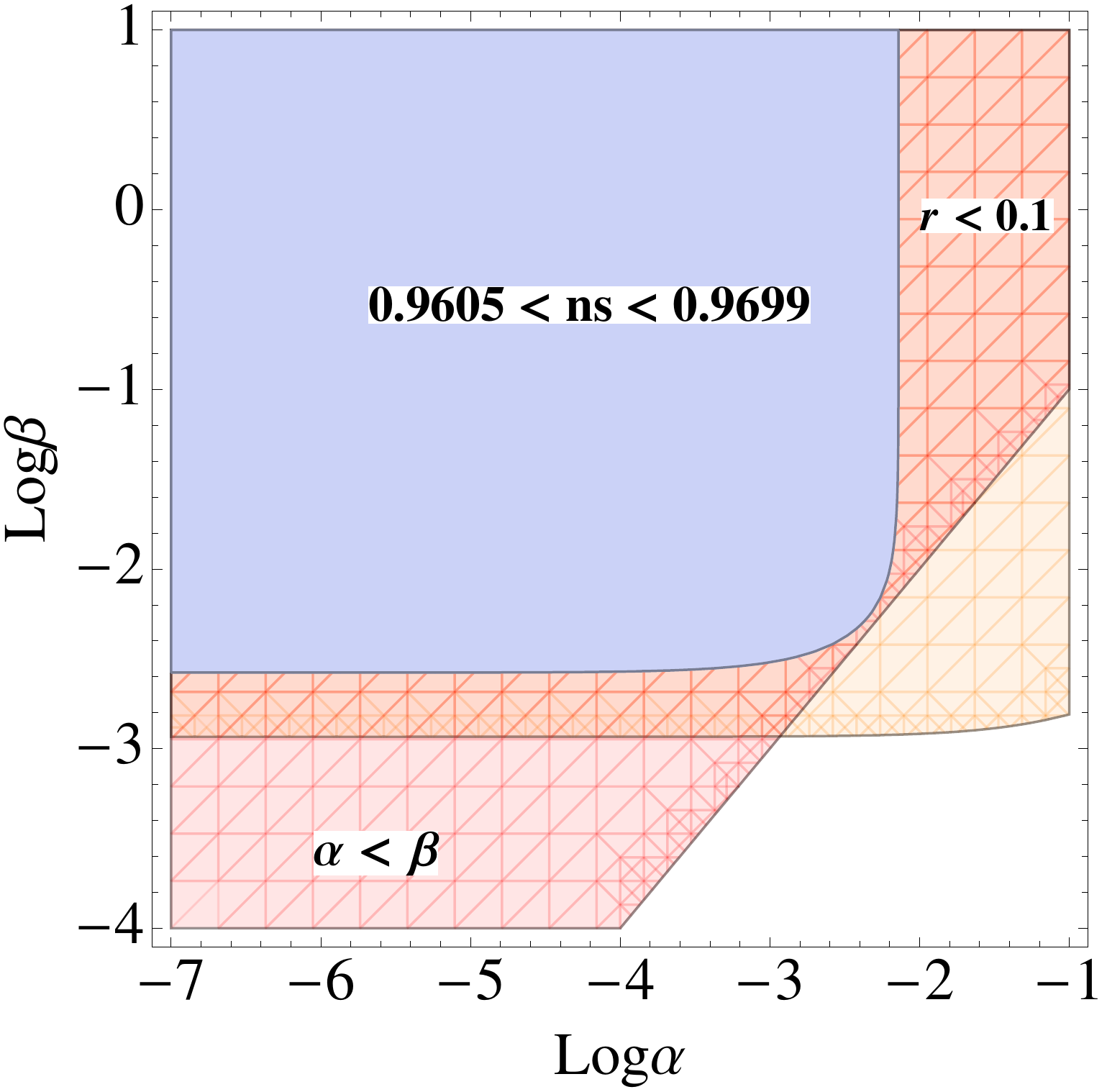}
        \caption{Constraints on the parameters $\alpha$ and $\beta$ for $\mathcal{N}_{\star}=60$.}
        \label{fig:N60}
    \end{minipage}
\end{figure}

\end{section}

\end{appendices}

\end{document}